\documentclass[a4paper,fleqn,usenatbib]{mnras}

\usepackage[T1]{fontenc}
\usepackage{ae,aecompl}

\usepackage{graphicx}	
\usepackage{amsmath}	
\usepackage{amssymb}	

\title[Zooming into local active galactic nuclei]{Zooming into local active galactic nuclei: The power of combining SDSS-IV MaNGA with higher resolution integral field unit observations}

\author[D.Wylezalek et al.]{Dominika Wylezalek$^{1}$\thanks{E-mail: dwylezalek@jhu.edu},
Allan Schnorr M\"{u}ller$^{2,3}$,
Nadia L. Zakamska$^{1,4}$,
\newauthor Thaisa Storchi-Bergmann$^{2,3}$,
Jenny E. Greene$^{5}$,
Francisco M\"{u}ller-S\'{a}nchez$^{6}$,
\newauthor Michael Kelly$^{1}$,
Guilin Liu$^{7}$,
David R. Law$^{8}$,
Jorge K. Barrera-Ballesteros$^{1}$,
\newauthor Rogemar A. Riffel$^{9,3}$,
Daniel Thomas$^{10}$
\\
$^{1}$Department of Physics \& Astronomy, Johns Hopkins University, Bloomberg Center, 3400 N. Charles St., Baltimore, MD 21218, USA\\
$^{2}$Departamento de Astronomia, Universidade Federal do Rio Grande do Sul, IF, CP 15051, 91501-970 Porto Alegre, RS, Brazil\\
$^{3}$Laborat\'orio Interinstitucional de e-Astronomia - LIneA, Rua Gal. Jos\'e Cristino 77, Rio de Janeiro, RJ - 20921-400, Brazil\\
$^{4}$Deborah Lunder and Alan Ezekowitz Founders' Circle Member, Institute for Advanced Study, Einstein Dr., Princeton, NJ 08540, USA\\
$^{5}$Department of Astrophysical Sciences, Princeton University, Princeton, NJ 08544, USA\\
$^{6}$Department of Astrophysical and Planetary Sciences, University of Colorado, Boulder, CO 80309, USA\\
$^{7}$CAS Key Laboratory for Research in Galaxies and Cosmology, Department of Astronomy, University of Science and Technology \\of China, Hefei, Anhui 230026, China\\
$^{8}$Space Telescope Science Institute, 3700 San Martin Drive, Baltimore, MD 21218, USA\\
$^{9}$Departamento de F\'isica, CCNE, Universidade Federal de Santa Maria, Av. Roraima, 1000 - 97105-900, Santa Maria, RS, Brazil \\
$^{10}$ Institute of Cosmology and Gravitation, University of Portsmouth, Dennis Sciama Building, Portsmouth, PO1 3FX, UK\\
}

\date{Accepted 01/26/2017}

\pubyear{2017}

\begin{document}
\label{firstpage}
\pagerange{\pageref{firstpage}--\pageref{lastpage}}
\maketitle

\begin{abstract}
Ionised gas outflows driven by active galactic nuclei (AGN) are ubiquitous in high luminosity AGN with outflow speeds apparently correlated with the total bolometric luminosity of the AGN. This empirical relation and theoretical work suggest that in the range $L_{\rm bol}\sim 10^{43-45}$~erg/s there must exist a threshold luminosity above which the AGN becomes powerful enough to launch winds that will be able to escape the galaxy potential. In this paper, we present pilot observations of two AGN in this transitional range that were taken with the Gemini North Multi-Object Spectrograph Integral Field Unit (IFU). Both sources have also previously been observed within the Sloan Digital Sky Survey-IV (SDSS) Mapping Nearby Galaxies at Apache Point Observatory (MaNGA) survey. While the MaNGA IFU maps probe the gas fields on galaxy-wide scales and show that some regions are dominated by AGN ionization, the new Gemini IFU data zoom into the centre with four times better spatial resolution. In the object with the lower $L_{\rm{bol}}$ we find evidence of a young or stalled biconical AGN-driven outflow where none was obvious at the MaNGA resolution. In the object with the higher $L_{\rm{bol}}$ we trace the large-scale biconical outflow into the nuclear region and connect the outflow from small to large scales. These observations suggest that AGN luminosity and galaxy potential are crucial in shaping wind launching and propagation in low-luminosity AGN. The transition from small and young outflows to galaxy-wide feedback can only be understood by combining large-scale IFU data that trace the galaxy velocity field with higher resolution, small scale IFU maps.

\end{abstract}

\begin{keywords}
galaxies: active -- galaxies: seyfert -- galaxies: kinematics and dynamics -- techniques: spectroscopic
 \end{keywords}


\section{Introduction}

The discovery of the tight relationship between black hole (BH) masses and the velocity dispersions and masses of their host bulges has shown that the active quasar phase of BH evolution has profound effects on galaxy evolution and that the enormous power of radiation and outflows from the BH and its accretion disk may be critical in limiting the maximum mass of galaxies in the universe \citep{Ferrarese_2005,Somerville_2008,Kormendy_2013}. However, constraining the power and reach of feedback processes exerted by luminous accreting BHs onto their hosts is still a field of active research in both simulations and observations \citep[e.g.][]{Brusa_2015b, Dugan_2016, Wylezalek_2016b, Zakamska_2016}

In low-luminosity active galactic nuclei (AGN, $L_{\rm{bol}} < 10^{43}$~erg/s), gas emission due to AGN photo-ionization may extend beyond several hundred parsecs, but the AGN-driven outflows are confined to the inner 50-200 pc \citep{Barbosa_2006, Barbosa_2009, Storchi-Bergmann_2009, Riffel_2011, Veilleux_2013, Lena_2015}. These flows carry only $< 0.1$\%$~L_{\rm{bol}}$ worth of power and rarely affect their large-scale environment. In contrast, extremely luminous quasars ($L_{\rm{bol}} > 10^{45}$~erg/s) often show signatures of high velocity, galaxy-wide outflows. Theoretical studies attribute this transition to the inertia of the interstellar medium: at a given galactic potential, an outflow driven by a low-luminosity AGN is not sufficiently powerful to accelerate the gas, so the gas inertia `quenches' the outflow before it can extend over galaxy scales \citep{Zubovas_2012}. 

Over the next several years, we endeavour to probe this transition using data from the Mapping Nearby Galaxies at Apache Point Observatory \citep[MaNGA;][]{Bundy_2015} survey. As part of Sloan Digital Sky Survey-IV \citep[SDSS-IV, ][]{Blanton_2017}, MaNGA is an integral field unit (IFU) survey targeting a statistically representative sample of 10,000 nearby ($0.01 < z < 0.15$) galaxies over the course of six years (survey start: July 2014) and is designed to explore the spatial dimension of galaxy evolution. The IFUs are made by grouping 19 to 127 fibres into hexagonal bundles, the number of fibres depending on the angular size of the galaxy. MaNGA uses the SDSS-III BOSS spectrograph \citep{Gunn_2006, Smee_2013, Drory_2015} and covers the wavelength range $3600-10,000~\AA$ with a velocity resolution of $\sim 60$~km/s. The spatial resolution is about $1-2$~kpc and most galaxies will be characterised out to 1.5 effective radii ($R_{e}$), a third of the galaxy sample even out to $2.5~R_{e}$ \citep{Law_2015, Yan_2016}.

The main MaNGA sample will contain some AGN ($\sim 300$ expected) but they span a very limited dynamical range in AGN luminosity typically reaching $L_{\rm{bol}} \sim 10^{43}$~erg/s. Therefore, in addition to the main MaNGA sample, a dedicated MaNGA-AGN program (PI: J.E.Greene) was awarded 120 MaNGA-IFU observations. These AGN cover a wide range of bolometric luminosities reaching $L_{\rm{bol}} \sim 10^{45}$~erg/s.

A key limitation of MaNGA is the large size of the fibres, 2\arcsec\ aperture (2.5\arcsec\ separation between fibre centres), which at $z\sim 0.05$ corresponds to $\sim 2$ kpc, although with dithering the effective sampling improves to $1.4$\arcsec. The wide-field coverage of MaNGA allows the mapping of global properties of the galaxies, but will not sample gas flows within the inner kiloparsec (kpc). Both models of AGN fuelling by gravitational instabilities \citep{Hopkins_2012} and observations to date demonstrate that at the luminosities of $10^{43}<L_{\rm bol}< 10^{45}$ erg/s the key AGN feeding and feedback processes occur within the inner kpc \citep{Barbosa_2014,Lena_2015}. Existing observations of AGN-driven winds in low-luminosity AGN where we expect the transition from circum-nuclear to galaxy-wide outflows have proceeded largely on an object-by-object basis, but establishing a comprehensive observational foundation of AGN feedback is only possible by surveying a large sample of AGN covering a wide parameter space. The MaNGA AGN sample is ideal to achieve these goals.  

In this paper, we present pilot IFU observations using the Gemini-North Multi-Object Spectrograph (GMOS-N) of two sources that have already been observed with MaNGA (before Oct 2015). At about 4 times better resolution compared to the MaNGA data, the goal of this paper is to map gas flows on scales $< 1$~kpc and to demonstrate the power of combining IFU data at different spatial scales and resolutions when exploring the launching and feeding mechanisms of low-luminosity AGN.  

The paper is organised as follows: Section 2 describes the target selection and both the MaNGA and GMOS observations. In Section 3 we present the analysis of the GMOS data, Section 4 explores the origin of the central gas flows and the added value of the GMOS data and in Section 5 we conclude. Throughout the paper we assume $H_0 = 71$ km s$^{-1}$ Mpc$^{-1}$, $\Omega_m = 0.27, \Omega_{\Lambda} = 0.73$. 

\section{Samples, observations and measurements}

\subsection{Target Selection}

\begin{table*}
\caption{Source Information}
\begin{center}
\begin{tabular}{lcccccc}
\hline\hline
Object Name & MaNGA ID & R.A. & Dec. & z & MaNGA plate-ID & MaNGA IFU-ID\\
\hline\vspace{0.05cm}
Blob Source & 1-166919 &    146.70910     &  43.423843      &0.0722 & 8459 & 3702\\
Cone Source & 1-137883 &  	  137.87476    &   45.468320       	  &   0.0268  &   8249 &  3704 \\
 \hline 
\end{tabular}
\end{center}
\label{observations}
\end{table*}

The GMOS observations were originally designed as pilot observations to explore the combination of large-scale MaNGA IFU data with smaller-scale, higher resolution GMOS observations. Only five bona-fide AGN from the ancillary MaNGA-AGN program had been observed at the time of the target selection. These ancillary AGN typically belong to the most luminous among all observed MaNGA-AGN. We therefore compiled a catalogue of weak AGN candidates in the main MaNGA sample. Selection of weak AGN in IFU data sets is a non-trivial task as we discuss in detail in an upcoming paper \citep{Wylezalek_2017} and in Section 4.1. We note that optical diagnostic diagrams are a crucial tool for this task.   

Diagnostic diagrams are constructed using a set of nebular emission lines and emission line ratios and can be used to distinguish between different ionization mechanisms of nebular gas. The most commonly used ones are the BPT diagrams \citep{Baldwin_1981, Veilleux_1987} using [NII]6584/H$\alpha$ versus [OIII]5007/H$\beta$ ([NII]-BPT diagram), [SII]6717,6731/H$\alpha$ versus [OIII]5007/H$\beta$ ([SII]-BPT) and [OI]6300/H$\alpha$ versus [OIII]5007/H$\beta$ ([OI]-BPT). A major advantage of the BPT diagnostic diagrams is that the required emission lines are relatively close in wavelength space such that usually all of them can be observed in one optical spectrum. Depending on ionization models several dividing lines have been developed such that the diagrams can be used to distinguish between different ionization mechanisms such as star formation, AGN or shocks. In this work, we use the dividing lines developed by \citet{Kewley_2001} and \citet{Kauffmann_2003} and summarised in \citet{Kewley_2006}. Specifically, the [SII]-BPT allows to distinguish between star-formation, AGN or `low ionisation nuclear emission line regions' \citep[LINER,][]{Heckman_1980} dominated emission line regions. The [NII]-BPT diagram allows to distinguish between star-formation, AGN/LINER, or composite dominated emission line regions. Because the [NII]-BPT diagram does not separate well between AGN or LINER-like emission, in the remaining part of the paper we refer to this emission as `AGN/LINER'-like to indicate that the selection based on the [NII]-BPT includes both classifications. LINER spectra show strong low ionization emission lines and characteristic line ratios which makes them easily identifiable in the [SII]-BPT diagram (but not in the [NII]-BPT). LINER-like emission line ratios can be produced in weak AGN and shocks, or in inactive galaxies due to photo-ionizing photons from hot evolved stars \citep[see e.g.][and references therein]{Ho_2008, Eracleous_2010, Singh_2013, Belfiore_2016}. 

\begin{figure*}
\centering
\vspace{1cm} 
\includegraphics[scale = 0.73, trim = 0.4cm 4cm 0cm 2cm, clip = true]{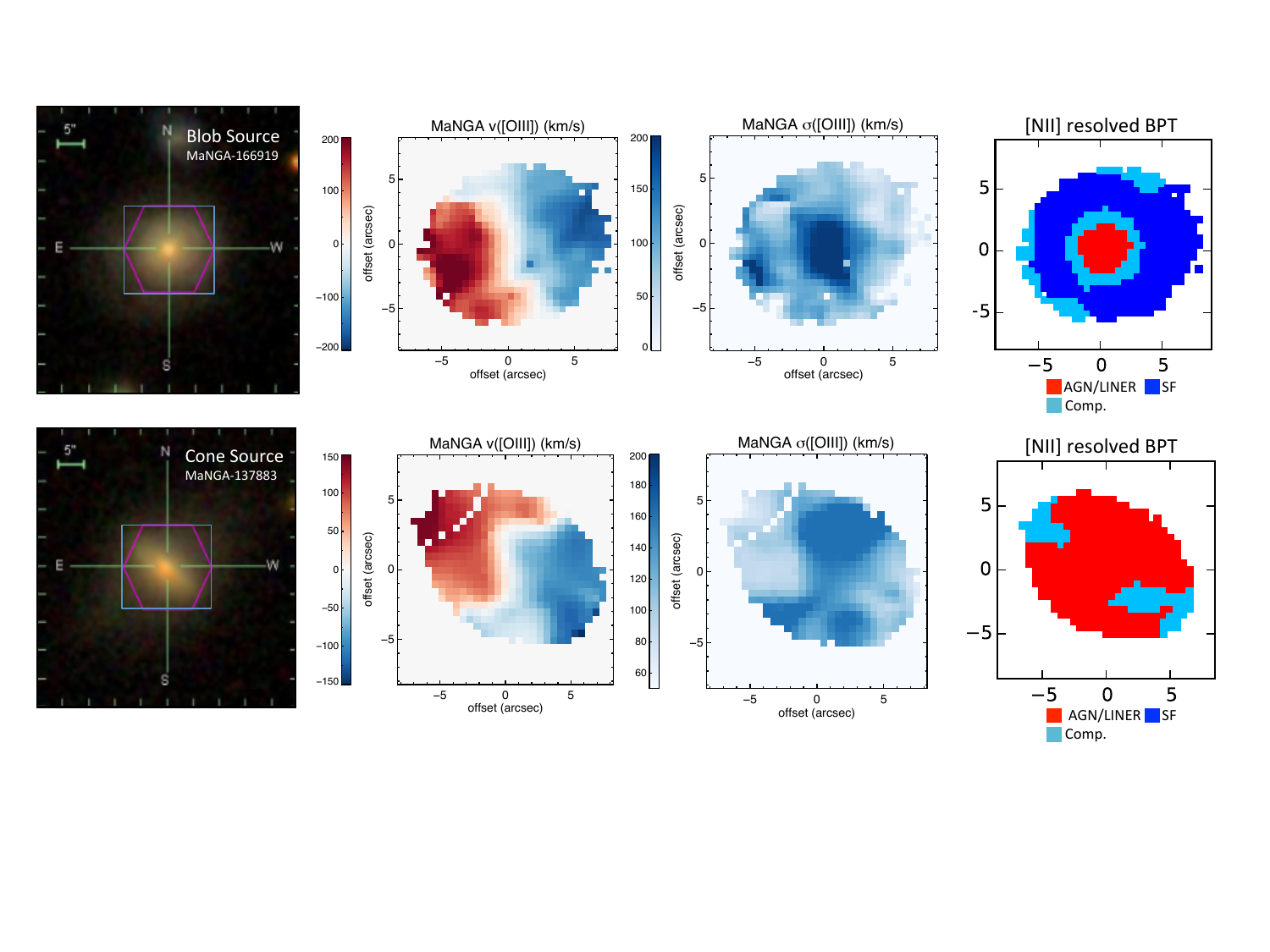}
\caption{MaNGA observations of the two sources analysed in this paper, with example data products produced by the MaNGA analysis pipeline and used for target selection. In the left panels, we show the SDSS composite image of the Blob (top) and Cone (bottom) AGN. The purple region shows the coverage with the MaNGA fibre bundle, while the blue boxes shows the $17.5\times17.5$~arcsec region covered by the MaNGA maps (right panels). The [OIII] velocity dispersion maps reveal areas of high velocity dispersion gas (dark blue regions). The resolved [NII] BPT diagram which determines which regions are dominated by AGN/LINER, star formation (SF) or both (Comp.), show that both objects have AGN/LINER dominated regions (red) where the [OIII] velocity dispersion is highest, indicative of a galactic wind. The Cone Source (bottom) shows two conical high ionization regions, while the Blob Source's (top) high ionization regions are round. With Gemini, we `zoom into' the central 3.5''$\times$5'' of these targets to trace the origin of the large-scale disturbed kinematics and to determine the power of AGN-driven winds}
\label{manga_data}
\end{figure*}

For the target selection for this work, we use a less conservative method and use the resolved [NII] BPT diagram to identify regions that show highly ionised gas (AGN and/or LINER-like emission, see red regions in the right panels in Figure \ref{manga_data}). We then visually inspect the morphology of these regions while simultaneously comparing with the morphology of high velocity dispersion ionised gas signatures traced by [OIII]. The final target selection was driven by trying to maximise both the range of bolometric luminosities as estimated from the [OIII] line \citep{Reyes_2008} and morphologies of ionised gas regions. The basic target information are listed in Table \ref{observations}.

The first source, MaNGA 1-166919, was chosen from the main MaNGA galaxy sample. It was selected as an AGN candidate based on its resolved [NII] BPT diagram which reveals a blob-like morphology of AGN/LINER-dominated emission with a radius of about 2~arcsec ($\sim 2.7$~kpc). This region is spatially coincident with a circular region of similar extent of high velocity dispersion ($\sigma$([OIII]) $> 200$~km/s) [OIII] emission (upper row in Figure \ref{manga_data}). Although originally not classified as an AGN in the SDSS classification based on the single-fibre spectrum, the spatially resolved MaNGA observations of MaNGA 1-166919 clearly show that a strongly ionizing source is residing in this galaxy. We hereafter dub the target the `Blob Source'. The galaxy stellar disk is observed close to face-on and its redshift is $z = 0.0722$.

From the sample of ancillary MaNGA AGN, we chose the only AGN identified by X-ray observations with the Burst Alert Telescope (BAT) aboard the \textit{Swift} satellite \citep{Ajello_2012}. This object, MaNGA 1-137883, at $z = 0.0268$, shows an intriguing cone-shaped morphology of high velocity [OIII] gas in MaNGA, suggestive of bi-polar outflows perpendicular to the plane of the stellar disk of the galaxy. This cone-shaped region coincides with AGN/LINER-like emission according to the [NII] resolved BPT diagram (lower row in Figure \ref{manga_data}). In the following we dub this source as `Cone Source'. 

The MaNGA fibre-bundle for both galaxies consists of 37 fibres with a diameter of 7 fibres, resulting in a bundle diameter of $17.5$~arcsec ($\sim 9$~kpc for the Cone Source, $\sim 24$~kpc for the Blob Source). The smallest spatial scale corresponding to the size of a single fibre in MaNGA is therefore $2.5$~arcsec ($\sim 1.3$~kpc for the Cone Source, $\sim 3.4$~kpc for the Blob Source). The reconstructed FWHM of the Blob and Cone Source are 2.8 and 2.4\arcsec\, respectively. The MaNGA Data Reduction Pipeline \citep[DRP, ][]{Law_2016} and Data Analysis Pipeline \citep[DAP, ][]{Westfall_2017} perform data reduction and output science-ready data cubes which provide refined data such as specific emission line fluxes and kinematics for each object surveyed. We use line and kinematic measurements from the DAP fourth internal MaNGA Product Launch (MPL-4) when exploring the MaNGA observations for the two objects presented in this work.

\subsection{GMOS observational setup}

The two objects were observed on December 2015 with the GMOS-IFU on Gemini-North (program ID: GN-2015B-FT-16, PI: D. Wylezalek). Both objects were observed in one-slit mode where the field-of-view is 3.5\arcsec$\times$5\arcsec, corresponding to a physical scale of $\sim 1.9 \times 2.7$~kpc for the Cone Source and $\sim 4.7 \times 6.8$~kpc for the Blob Source, for a total exposure time of 1 hour, respectively. We used the B600-G5307 grating covering rest-frame wavelengths of $\sim 4000 - 7000~\AA$. This wavelength range covers most of major optical emission lines including [OIII]4363,4959,5007, [NII]6548,6583, H$\alpha$ and H$\beta$. For both objects, we took six science exposures of 600~s each, and dithered both in the spatial (by 0.2\arcsec) and spectral direction (by 100\AA) in order to minimise the impact of bad pixels or cosmic rays and to cover the spectral gap of $\sim 100\AA\ $ between the spectrographs. Seeing was $\sim 0.9$\arcsec, improving on the spatial resolution of MaNGA observations by a factor of $\sim 3$.

\subsection{Data Reduction}

The data reduction was performed using specific tasks developed for GMOS data in the gemini.gmos package as well as generic tasks in iraf. The reduction process comprised bias subtraction, flat-fielding, trimming, wavelength calibration, sky subtraction, relative flux calibration, building of the data cubes at a sampling of $0.1$\arcsec $\times 0.1$\arcsec\, and finally the alignment and combination of the data cubes.

\subsection{Spectral Fitting}

We develop a customised fitting procedure to extract the emission line fluxes and kinematics from the GMOS data. In every spatial element (spaxel) we subtract a linear continuum which we estimate from the average flux level red- and blue-ward of the emission lines to be fitted. In case of line multiplets or where emission lines lie very close in wavelength such as the [OIII] doublet at 5007$\AA$, 4959$\AA$ and H$_{\beta}$ at 4861$\AA$ we estimate the continuum flux level red- and blue-ward of such line complexes. We then use non-parametric measurements that do not strongly depend on a specific fitting procedure to determine amplitudes, centroid velocities and emission line widths. We follow the measurement strategy presented in e.g. \citet{Zakamska_2014} and \citet{Liu_2013b}.

Briefly, each profile is first fitted with multiple Gaussian components to determine the overall velocity profiles. Despite the complexity of the velocity structure, most emission lines can be well fit with two or at most three Gaussian profiles. For subsequent analysis we use the fit that minimises both the number of Gaussian components and $\chi^2$ of the fit.  

\begin{figure}
\centering
\vspace{1cm} 
\includegraphics[scale = 0.45, trim = 0cm 0.7cm 0cm 2.5cm, clip= true]{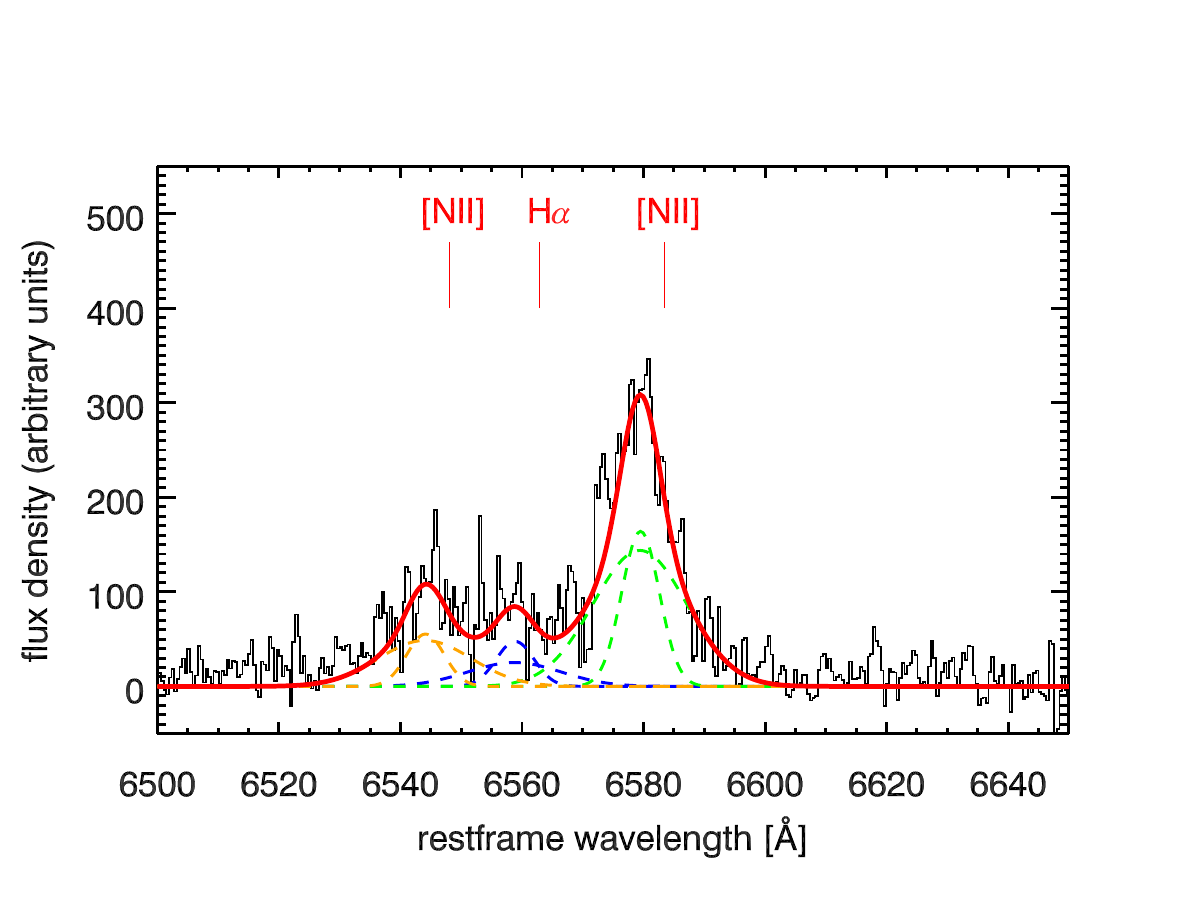}
\caption{Example spectrum (black) and fit (red solid line) to the H$\alpha$ and [NII] lines in one of the spaxels of the Blob Source's GMOS data cube. All three lines were fitted with the same kinematic model with $W_{80} = 670$~km/s. The lines are blue-shifted with respect to the rest-frame wavelength (shown by the red vertical bars) by $-60$~km/s. The blue, green and orange dashed lines show the individual Gaussian components to the H$\alpha$ and [NII] lines, respectively.}
\label{spectrum_ex}
\end{figure}

The cumulative flux as a function of velocity is then:
\begin{equation}
\Phi(v) = \int_{-\infty}^{v} F_{v}(v') dv'
\end{equation}
such that the total line flux is given by $\Phi(\infty)$. 
In practice, we use the $[-10000, 10000]$~km~s$^{-1}$ range in the rest-frame as our maximal interval of integration. For each spectrum, this definition is used to compute the line-of-sight velocity, represented by the median velocity $v_{med}$, the velocity that bisects the total area underneath the emission-line profile, so that $\Phi(v) = 0.5\ \Phi(\infty)$. Since stellar absorption lines are too faint in the GMOS spectra we adopt the spectroscopic redshifts from the NASA Sloan Atlas (NSA) catalogues that are based on the single-fibre measurements to correct the spectra to the rest-frame of the galaxy. We measure the line widths using the non-parametric measurement $W_{80}$ which corresponds to the velocity width that encloses 80\% of the total flux. For a purely Gaussian profile, $W_{80}$ is closely related to the FWHM with $W_{80} = 1.088 \times$ FWHM, but the non-parametric velocity width measurements are more sensitive to the weak broad bases of non-Gaussian emission line profiles \citep{Liu_2013b}. The kinematics (i.e. $W_{80}$ and $v_{med}$) of different lines within a complex, such as [OIII]+H$\beta$ or [NII]+H$\alpha$ are assumed to be the same. Additionally, we fix the ratio [OIII]5007$\AA$/[OIII]4959$\AA$ and [NII]6583$\AA$/[NII]6548$\AA$ to their quantum values of 2.98 and 2.95, respectively. In Figure \ref{spectrum_ex} we show an example fit to the H$\alpha$ and [NII] emission lines in a central spaxel in the GMOS data cube of the Blob Source. 

H$\alpha$ and H$\beta$ emission lines lie on top of stellar absorption features. To test the sensitivity of our line measurements to stellar continuum subtraction, we convert MaNGA spectra from vacuum to air wavelengths \citep{Morton_1991} and then average stellar continuum contributions as measured by DAP over all spaxels to construct a high signal-to-noise host continuum. We subtract this scaled continuum from individual GMOS spectra, then perform an additional linear continuum subtraction to account for small spectrophotometric deviations of GMOS data and refit emission lines. We do not find any significant deviations in line fitting parameters and therefore base our analysis on the simple linear continuum fitting as described above. We show the resulting kinematic maps in Figures \ref{GMOS_blob}, \ref{GMOS_blob_Nii}, \ref{GMOS_cone_vel} and \ref{GMOS_cone_disp}.

\section{Ionised Gas Morphologies and Kinematics}

The two AGN in this work were (amongst other criteria) selected based on the prevalence, morphologies and kinematics of the ionised gas emission. In both objects the high velocity dispersion region of the [OIII] ionised gas, the dark blue regions in Figure \ref{manga_data}, are spatially coincident with the AGN/LINER-dominated spaxels in the resolved [NII]-BPT maps. The main advantage of the GMOS observations is that it allows us to resolve the inner few kiloparsecs of this AGN/LINER-dominated emission and probe their origin. In the following subsection, we investigate the kinematics and morphologies of the ionised gas using the better resolution GMOS data. 

\subsection{Blob Source}

\begin{figure*}
\centering
\textbf{Blob Source}\par\medskip
\includegraphics[scale = 0.35, trim = 1cm 3.1cm 6cm 12cm, clip = true]{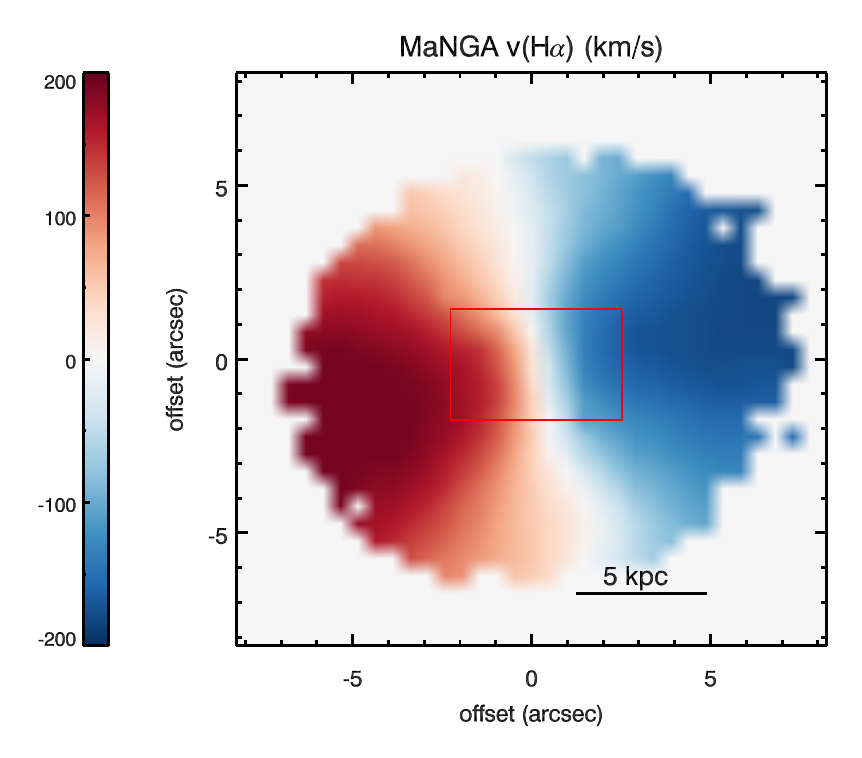}
\includegraphics[scale = 0.35, trim = 1cm 3.1cm 6cm 12cm, clip = true]{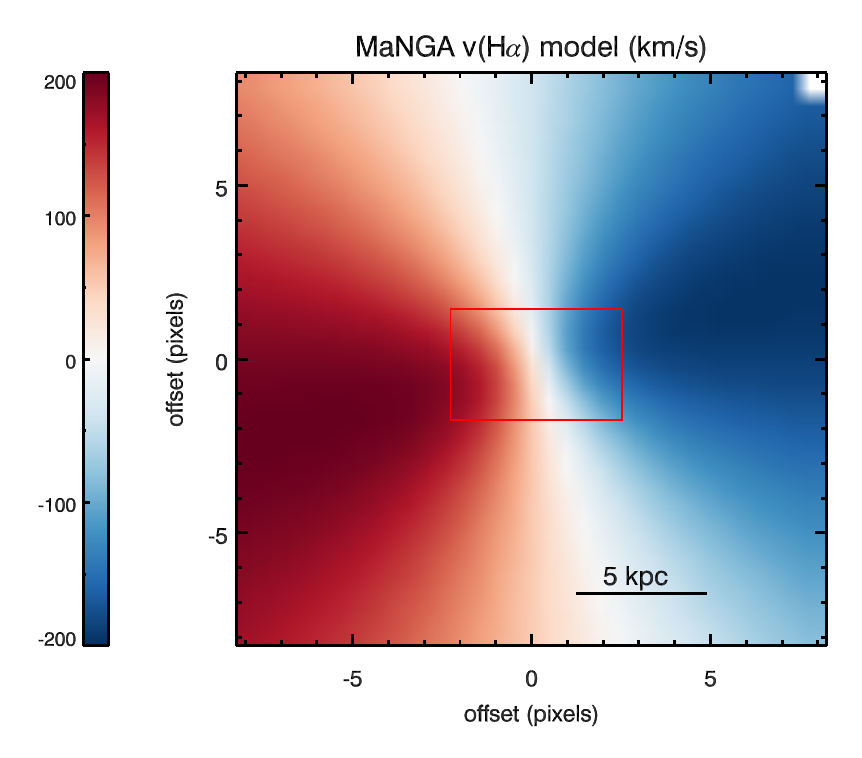}
\includegraphics[scale = 0.35, trim = 1cm 3.1cm 3cm 12cm, clip = true]{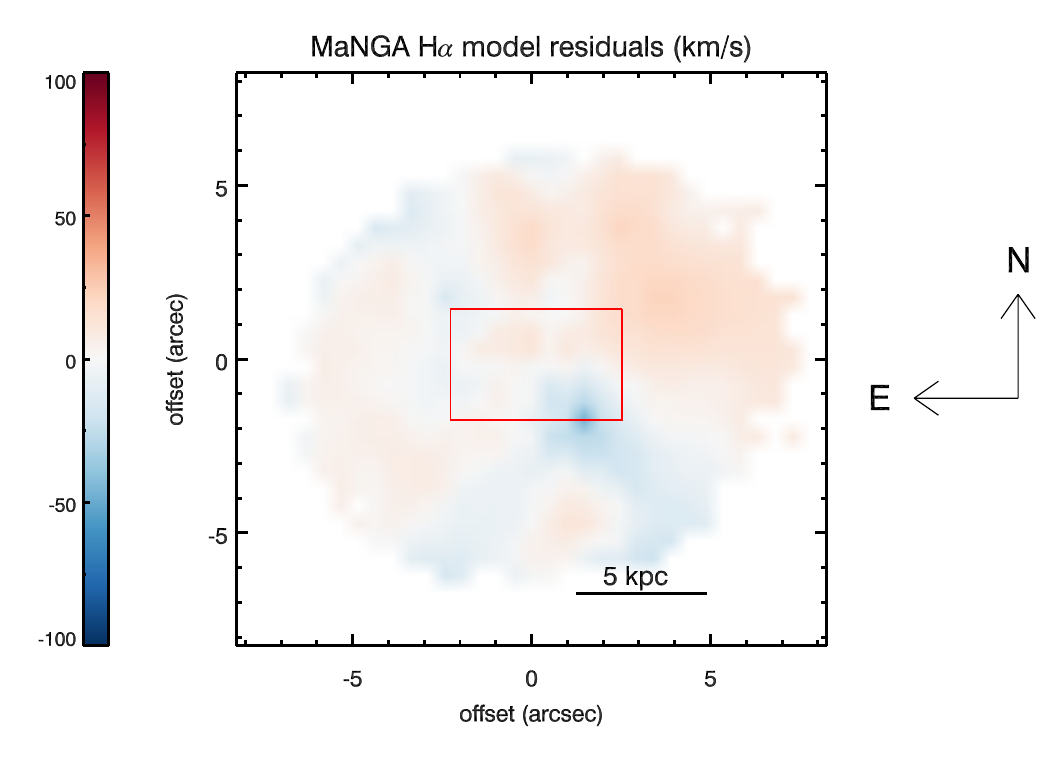}
\includegraphics[scale = 0.35, trim = 1cm 3.1cm 6cm 14cm, clip = true]{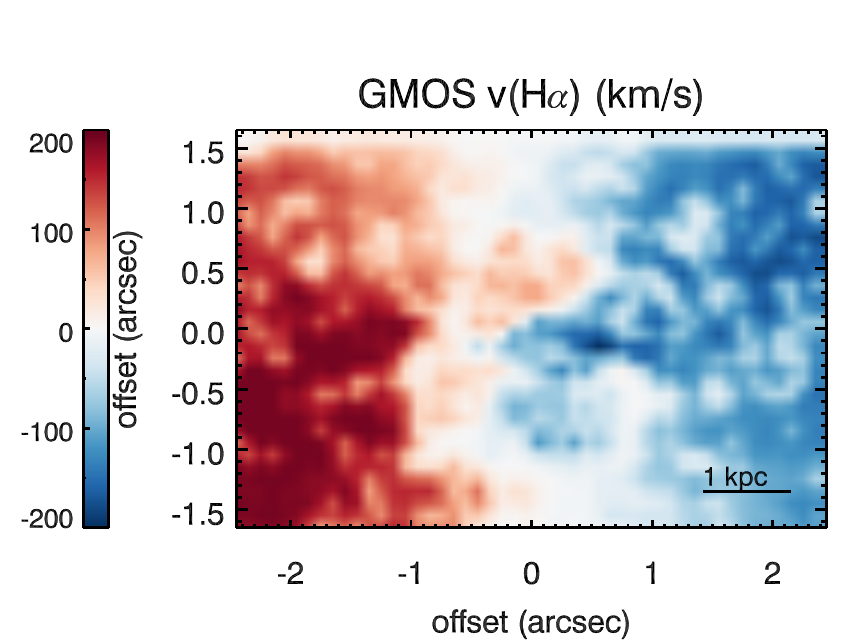}
\includegraphics[scale = 0.35, trim = 1cm 3.1cm 6cm 14cm, clip = true]{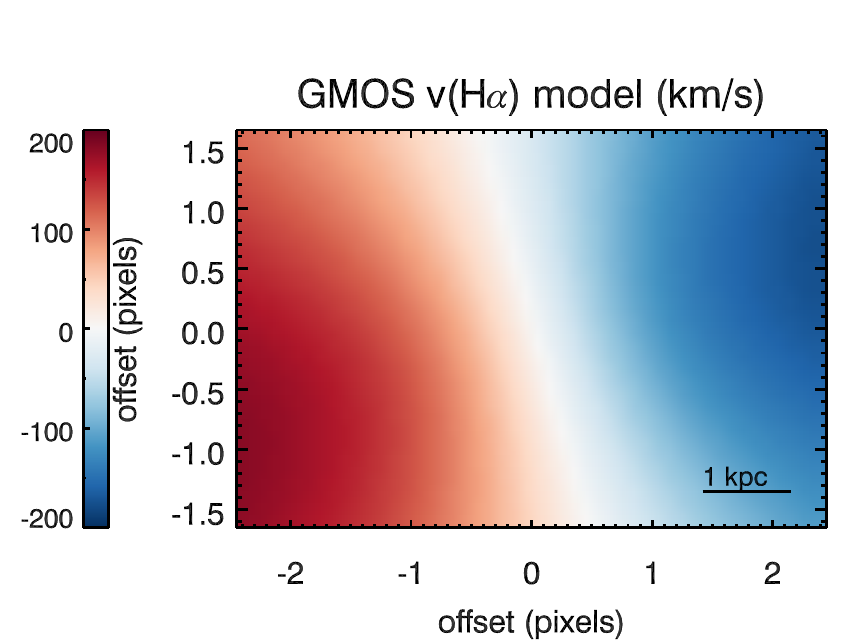}
\includegraphics[scale = 0.35, trim = 1cm 3.1cm 3cm 14cm, clip = true]{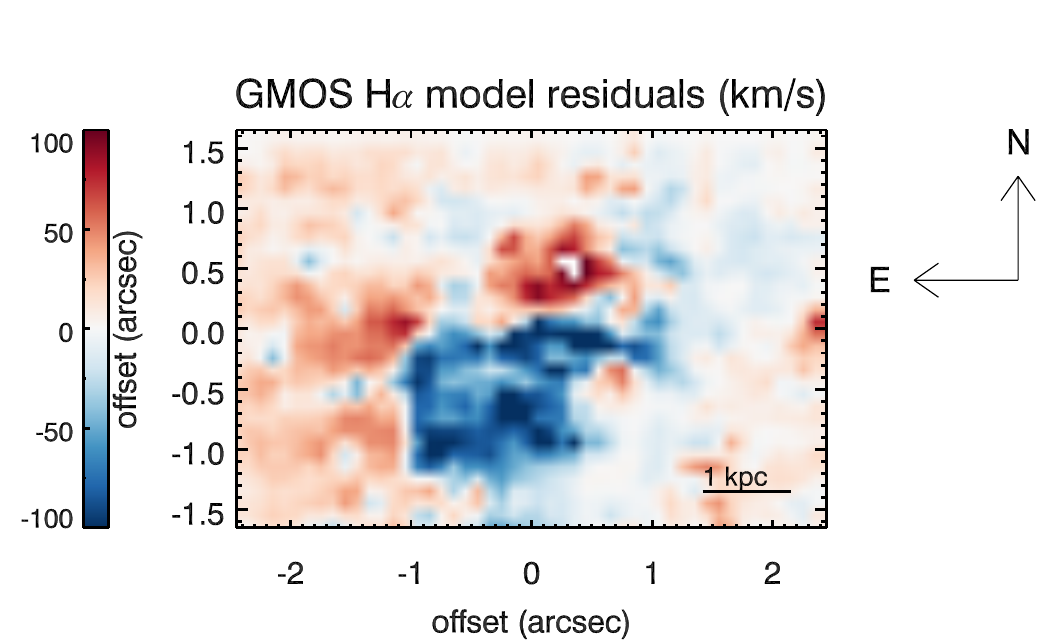}
\caption{H$\alpha$ velocity map as observed with MaNGA and with GMOS (left panels). The GMOS observations zoom into the central part of the Blob Source, which is indicated by the red box in the MaNGA maps. The middle panels show the rotational model to the H$\alpha$ velocity fields and the right panels the residuals (data$-$model). While the model describes the MaNGA velocity field very well (residuals are negligible), the blue- and redshifted residuals from GMOS reveal the bipolar nature of a kinematically decoupled structure in the centre of the source.}
\label{GMOS_blob}
\end{figure*}

The MaNGA data of the Blob Source reveals a spherically symmetric region with a radius of $\sim 3$~arcsec of AGN/LINER-dominated emission (Figure \ref{manga_data}). Since the [OIII] and H$_\beta$ emission lines are not detected in the GMOS data due to a low signal-to-noise ratio, we use the H$\alpha$ line to identify peculiarities in the gas velocity field. The MaNGA H$\alpha$ velocity field shows ordered rotation of the gaseous disk of the host galaxy, which is the dominant effect even though the galaxy is viewed close to face-on (upper left panel in Figure \ref{GMOS_blob}). In Figure \ref{GMOS_blob} (lower left panel) we also show the central region of the H$_\alpha$ gas velocity field as observed by GMOS. With the higher resolution data, we clearly see deviations from ordered rotation in the gas field near the nucleus.

In order to isolate these non-circular motions of the ionised gas and quantify these deviations, we need to subtract a rotating disk component from the gas velocity field. As non-circular motions in the gas velocity field can be significant, fitting an unconstrained model is misleading, as the kinematic centre and the position angle of the line of nodes are model parameters significantly influenced by non-circular motions. Additionally, the stellar and gas disk components do not necessarily match, as the stellar component can rotate more slowly than the gas \citep{van_der_Kruit_1986, Bottema_1987, Noordermeer_2008}. Thus, we utilise the large-scale MaNGA rest-frame stellar velocity field and determine the kinematic centre and the position angle of the line of nodes from a fit to the MaNGA stellar velocity field. We model the stellar velocity field assuming a spherical potential with pure circular motions, with the observed radial velocity at a position ($R,\psi$) in the plane of the sky given by \citet{Bertola_1991}:

\begin{equation}
V=\frac{AR\cos(\psi-\psi_{0})\sin(\theta)\cos^{p}\theta}{\{R^{2}[\sin^{2}(\psi-\psi_{0})+\cos^{2}\theta \cos^{2}(\psi-\psi_{0})]+c^{2}\cos^{2}\theta \}^{p/2}} 
\end{equation}
where $\theta$ is the inclination of the disk (with $\theta$\,=\,0 for a face-on disk), $\psi_{0}$ is the position angle of the line of nodes, $R$ is the radius, $A$ is the amplitude of the rotation curve, $c$ is a concentration parameter regulating the compactness of the region with a strong velocity gradient and $p$ regulates the inclination of the flat portion of the velocity curve. We perform a Levenberg-Marquardt least-squares minimization to determine the best fitting parameters.
We then fit this rotating disk model to the large-scale H$\alpha$ rest-frame velocity field probed by MaNGA, assuming the gas and stars have the same orientation and kinematic centre. The amplitude of the rotation curve, the concentration parameter and the inclination of the flat portion of the velocity curve are free parameters in the fit.

The resulting fits are shown in the middle panels in Figure \ref{GMOS_blob} and the residuals (i.e. H$\alpha_{\rm{measured}}$-H$\alpha_{\rm{model}}$) in the right panels in Figure \ref{GMOS_blob}. While the fit to the MaNGA H$\alpha$ velocity field describes the data very well with negligible residuals, the GMOS H$\alpha$ residual maps shows a clear gradient in residual velocities with a redshifted Northern region and a blueshifted Southern region. In wide-angle winds, the velocity dispersion usually reflects the typical bulk velocities of the gas, while the observed median velocities are more sensitive to projection and geometric orientation effects \citep[see also][]{Liu_2013b}. These kinds of signatures are therefore indicative of a bipolar structure that is kinematically decoupled from the gas disk velocity field.

However, such model residuals can be sensitive to changes in the fitted position angle of the major axis. To quantify the significance of the detection of a kinematically decoupled component, we conduct the following test: We add random noise on the order of the measurement uncertainties (typically 10\%, i.e. $15-20$~km/s) to the velocity field, and repeat the model fit 100 times to obtain the uncertainties on the model parameters. The position angle of the major axis is well constrained, with $1\sigma$ uncertainties of only 0.51 degrees. Varying the position angle by as much as $\pm 2\sigma$ has little effect on the residuals and on the properties of the detected kinematically decoupled component.

\citet{Fischer_2016} recently showed that the ionised gas kinematics in the nearby ($z = 0.017$) Seyfert 2 galaxy Mrk 573 largely agree with the stellar kinematics and are consistent with rotation. They attribute the morphology and spatial extent of the detected ionised emission (traced by [OIII] in their case) to gas that is in rotation in the galactic disk and only gets passively ionised, i.e. `illuminated', when the central AGN turns on. The situation in our Blob Source is clearly different from Mrk 573, in that we robustly detect the elongated kinematic structure in the H$\alpha$ residual map. Our modelling shows that the nuclear ionised gas is kinematically decoupled from the disk rotation and appears to be undergoing a bipolar outflow instead.

Motivated by these findings, we proceed with comparing the GMOS H$\alpha$ velocity maps with models comprising biconical outflows superimposed on the disk rotation derived from MaNGA data. Details of the modelling can be found in \citet{Muller-Sanchez_2011}. Briefly, the bi-cone model consists of two symmetrical hollow cones having interior and exterior walls with apexes coincident with a central point source. The simplest velocity law which reproduces the line-of-sight velocity fields of nearby AGN is radial linear acceleration followed by radial linear deceleration \citep[see also][]{Crenshaw_2010, Fischer_2013}. We find a maximum extent of the two cones combined to be $\sim 3$~kpc with a maximum de-projected outflow velocity of $v_{\rm{max}} \sim 260$~km/s, confirming our observation above. The turnover radius $r_t$, at which $v_{\rm{max}}$ is observed, is measured to be 800~pc. We estimate the mass outflow rate $\dot{M}_{out}$ using the method described in \cite{Muller-Sanchez_2011} and assuming a gas density $n_e  = 100$~cm$^-3$ and a filling factor $f  = 0.01$, typical values of the narrow-line region at $r\sim1$~kpc. The resulting $\dot{M}_{out}$ is $66$~M$_{\sun}$yr$^-1$ and the total kinetic power is $\dot{E}_{\rm{kin}} = 1.4\times10^{42}$~erg s$^{-1}$. Due to large uncertainties in $f$ and $n_e$ these estimates are probably accurate to no better than about $\pm0.5$~dex.

Considering all the described evidence strongly suggests that the kinematically distinct H$\alpha$ component is in fact a biconical outflow propagating at $\sim 200$~km/s. Because of its small size of $\sim 3$~kpc at its widest extent, this component had not been resolved in the lower spatial resolution MaNGA observations.

\begin{figure}
\centering
\textbf{Blob Source}\par\medskip
\includegraphics[scale = 0.28, trim = 1.5cm 3.1cm 6cm 12cm, clip = true]{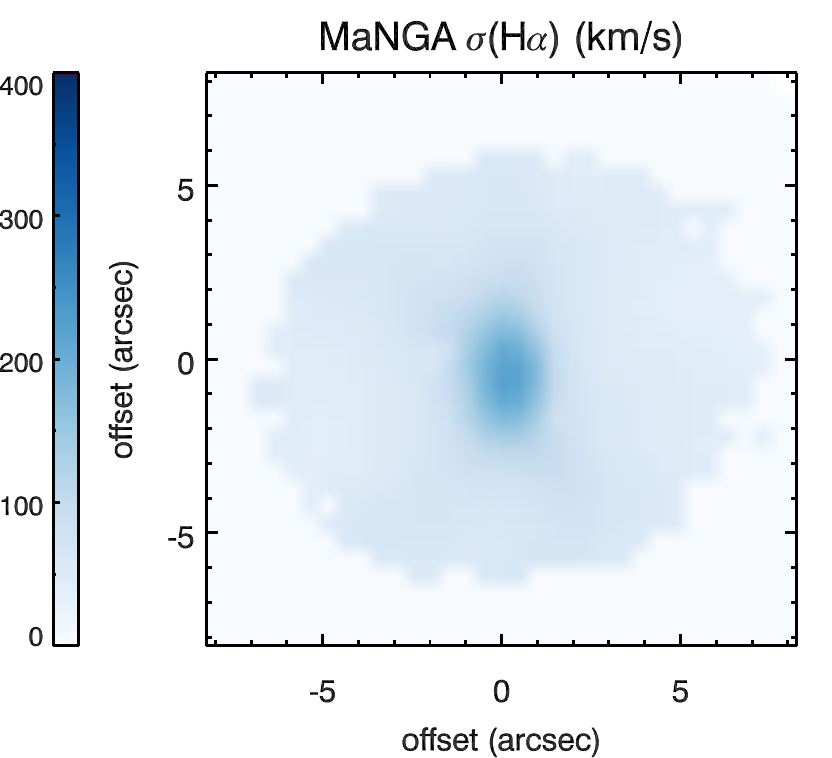}
\includegraphics[scale = 0.28, trim = 1.4cm 3.1cm 6cm 12cm, clip = true]{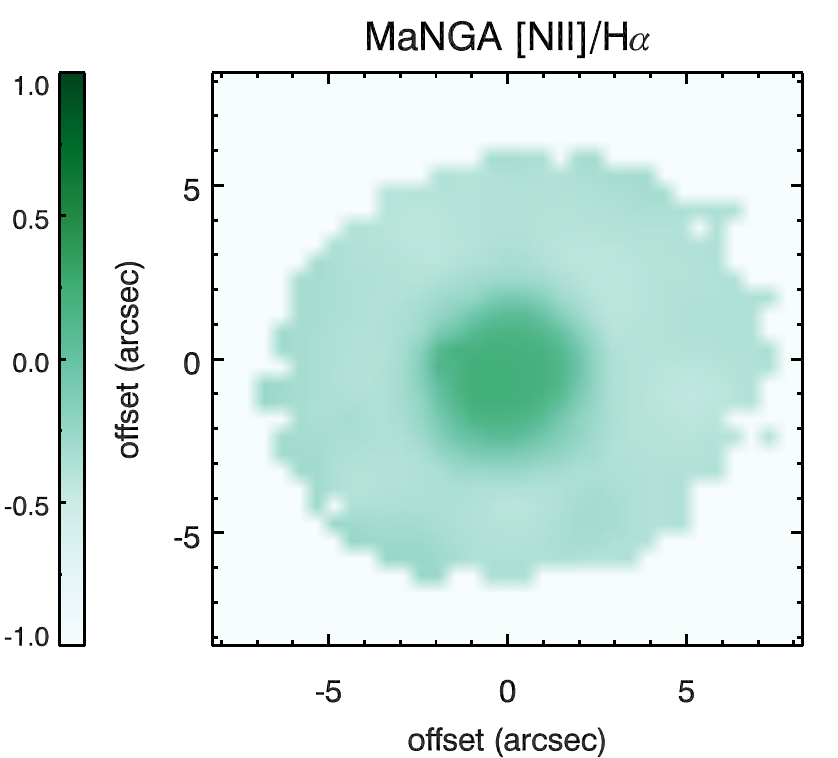}
\includegraphics[scale = 0.28, trim = 1.5cm 3.1cm 6cm 14cm, clip = true]{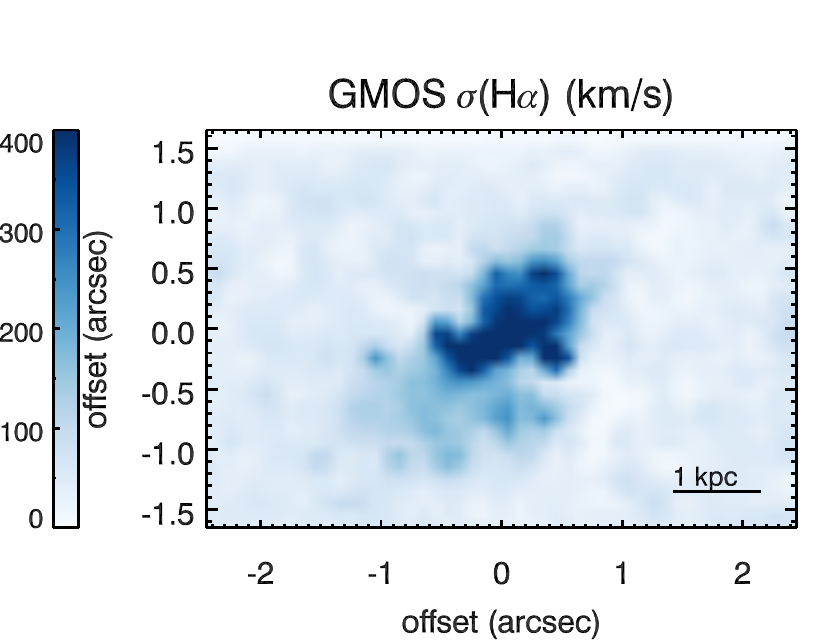}
\includegraphics[scale = 0.28 , trim = 1.4cm 3.1cm 6cm 14cm, clip = true]{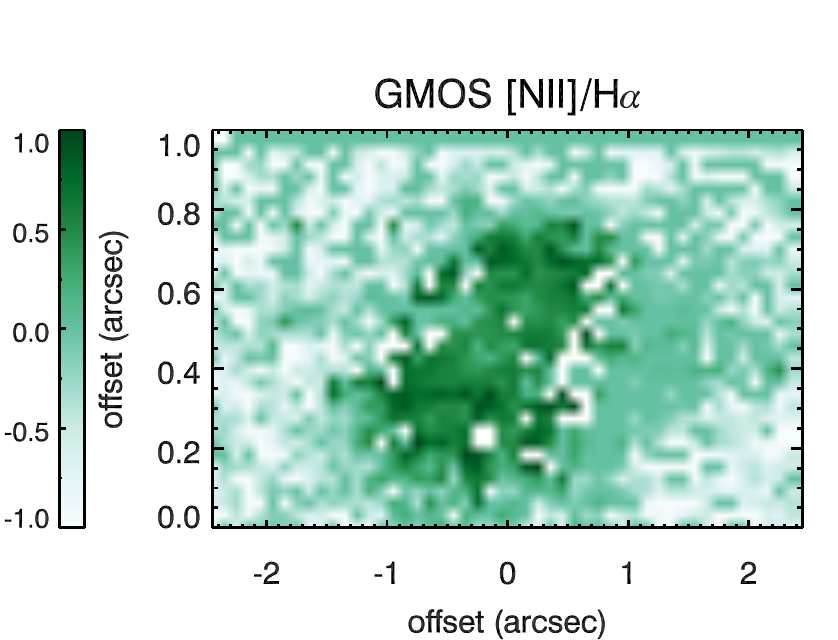}
\caption{The kinematically decoupled structure that we show in Figure \ref{GMOS_blob} spatially coincides with a region of high H$_{\alpha}$ velocity dispersion (left panels) and high [NII]/H$\alpha$ ratios (right panels), an indication that this structure is either shock- or AGN-ionised and not due to nuclear star formation. The size of the structure is only about 2~kpc across, was not detected in the MaNGA data and is potentially a young or stalled outflow. We note that the GMOS $\sigma(H_{\alpha})$ corresponds to $0.4*W_{80}$($H_{\alpha}$).}
\label{GMOS_blob_Nii}
\end{figure}

In Figure \ref{GMOS_blob_Nii} we additionally show the H$_{\alpha}$ velocity dispersion maps and [NII]/H$\alpha$ maps as observed by both MaNGA and GMOS. In the GMOS map we find a region with $\sigma_{H\alpha}$ of $\sim 400$~km/s that is observed at the same position as the distinct biconical H$\alpha$ component described above. The position angle of this structure is about -45 degrees and its spatial extent is $\sim 3$~kpc. High [NII]/H$\alpha$ ratios are indicative of ionization through either (SF-driven or AGN-driven) shocks or AGN photoionization. In Section 4.1, we explore these signatures further.

\subsection{Cone Source}

\begin{figure*}
\centering 
\textbf{Cone Source}\par\medskip
\includegraphics[scale = 0.34, trim = 1cm 3.1cm 6cm 11cm, clip = true]{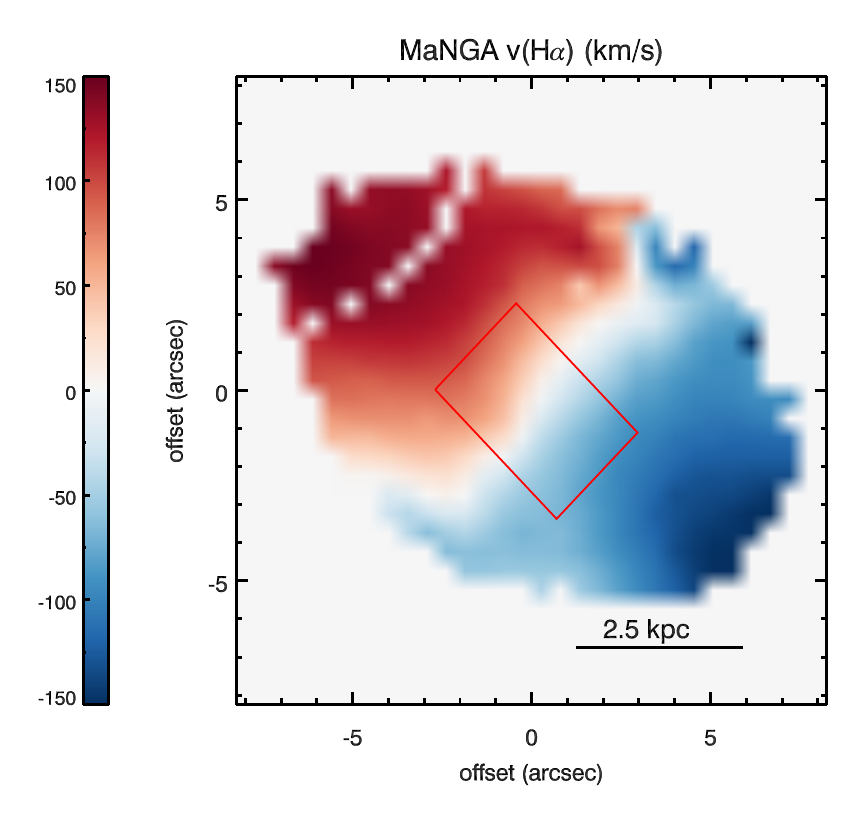}
\includegraphics[scale = 0.34, trim = 1cm 3.1cm 6cm 11cm, clip = true]{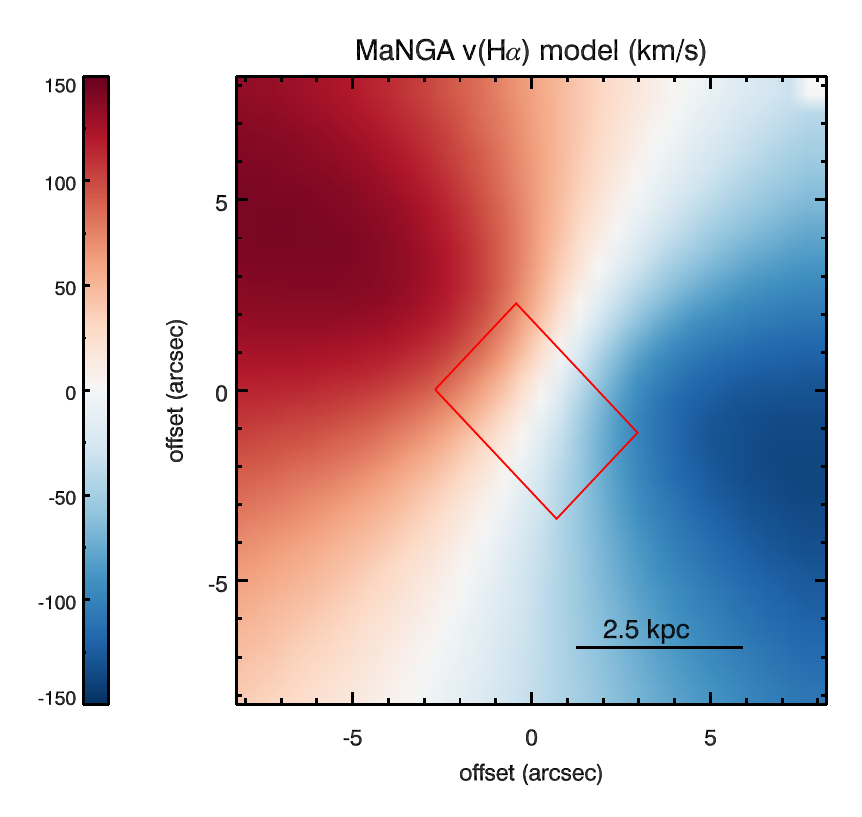} 
\includegraphics[scale = 0.34, trim = 1cm 3.1cm 3cm 11cm, clip = true]{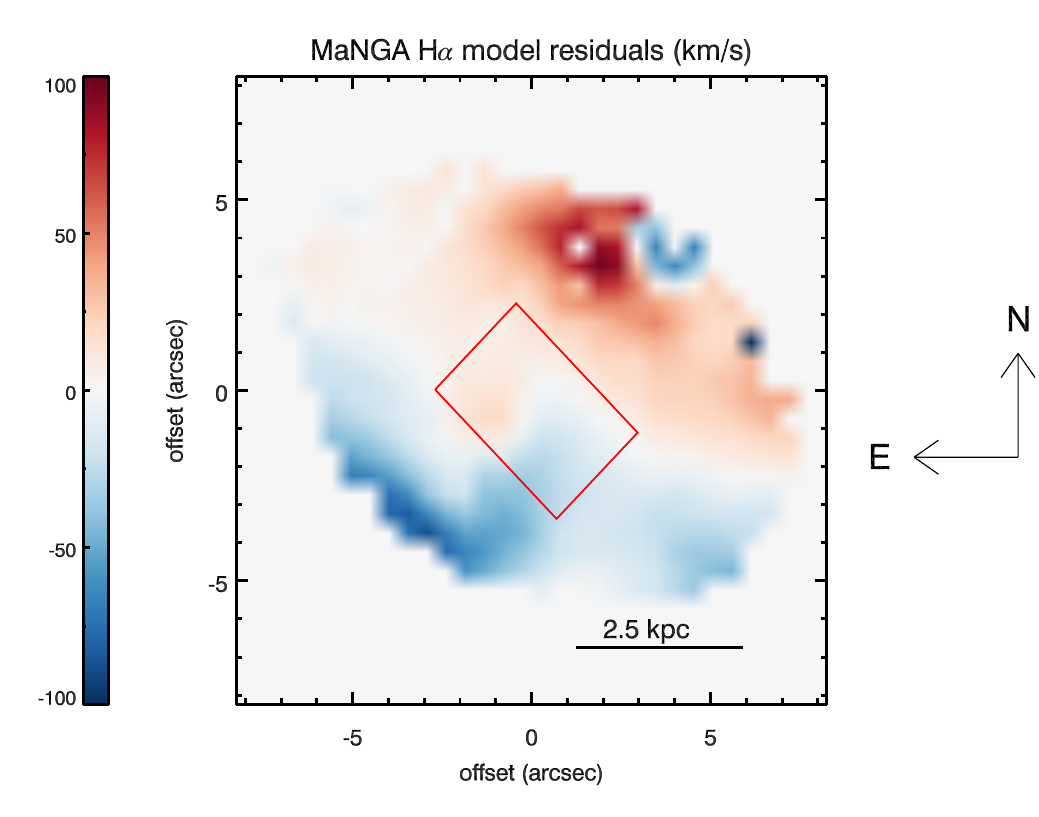}
\includegraphics[scale = 0.34, trim = 1cm 3.1cm 6cm 14cm, clip = true]{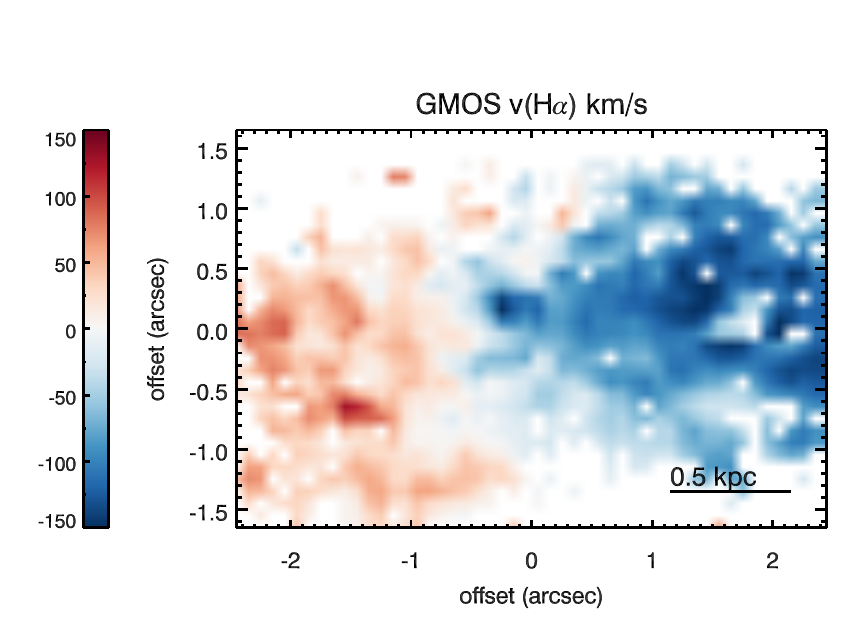}
\includegraphics[scale = 0.34, trim = 1cm 3.1cm 6cm 14cm, clip = true]{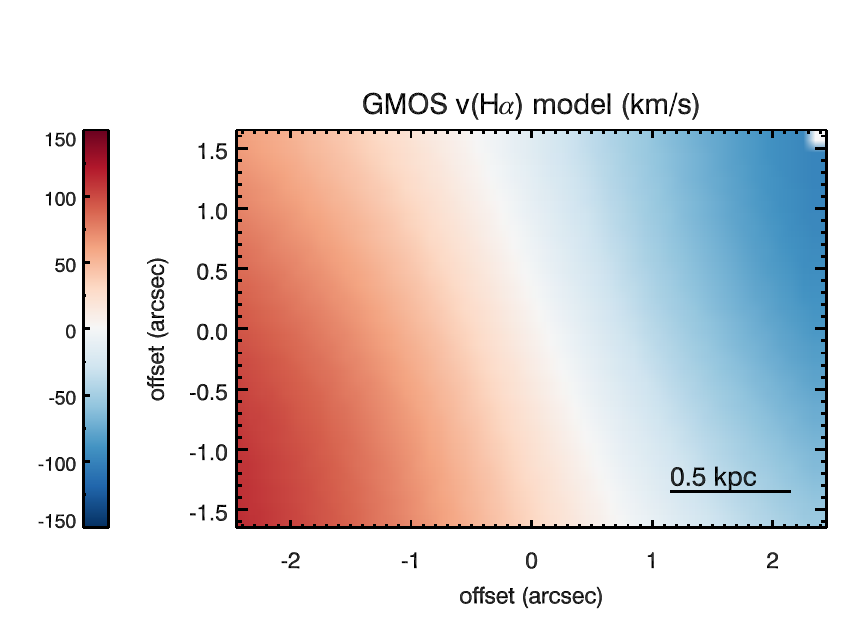}
\includegraphics[scale = 0.34, trim = 1cm 3.1cm 3cm 14cm, clip = true]{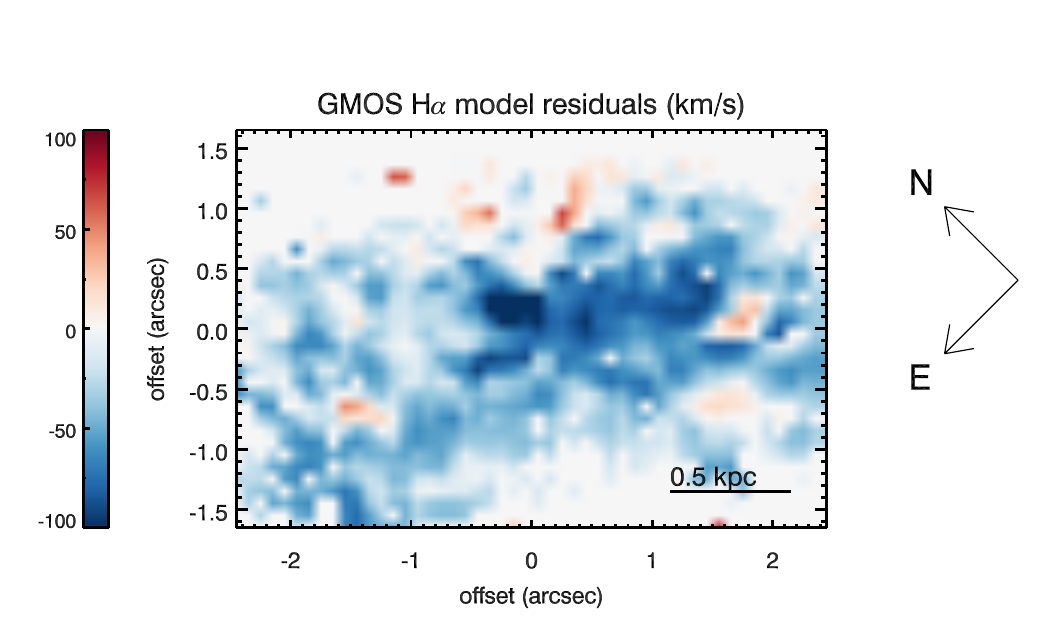}
\caption{H$\alpha$ velocity map as observed with MaNGA and with GMOS (left panels). The GMOS observations zoom into the central part of the Cone Source, which is indicated by the red box in the MaNGA maps. The middle panels show the rotational model to the H$\alpha$ velocity fields and the right panels the residuals (data$-$model). The MaNGA residual map reveals prominent residuals that are co-spatial with regions of high velocity ionised gas (see Figure \ref{GMOS_cone_disp}). The outflow in the centre of the galaxy seems warped and only blue-shifted residuals are apparent in the GMOS H$\alpha$ residual map.}
\label{GMOS_cone_vel}
\end{figure*}

In contrast to the Blob Source, the MaNGA data for the Cone Source already showed a biconical morphology of fast $\sigma \sim 200$~km/s ionised gas as traced by the [OIII] emission line. By analogy to Figure \ref{GMOS_blob}, we show the H$\alpha$ velocity field as measured by both MaNGA and GMOS in Figure \ref{GMOS_cone_vel}. We follow the same strategy as described above and fit rotational models to the velocity fields that are shown in the middle panels in Figure \ref{GMOS_cone_vel} and compute the residuals (right panels in Figure \ref{GMOS_cone_vel}). We identify prominent residuals of bi-polar nature in the MaNGA residual map. We re-run the fitting procedure after having added random noise to the velocity maps to estimate the uncertainties on the fitting parameters, specifically the on the position angle of the major axis, and significance of the residuals. The $1\sigma$ uncertainty on the position angle of the major axis is only $\sim 0.7$~degrees. The detection of the prominent residuals is therefore robust and indicative of a gas kinematic component that is decoupled from the disk velocity field.

\begin{figure}
\centering
\textbf{Cone Source}\par\medskip
\includegraphics[scale = 0.28, trim = 1.3cm 3.1cm 6cm 11cm, clip = true]{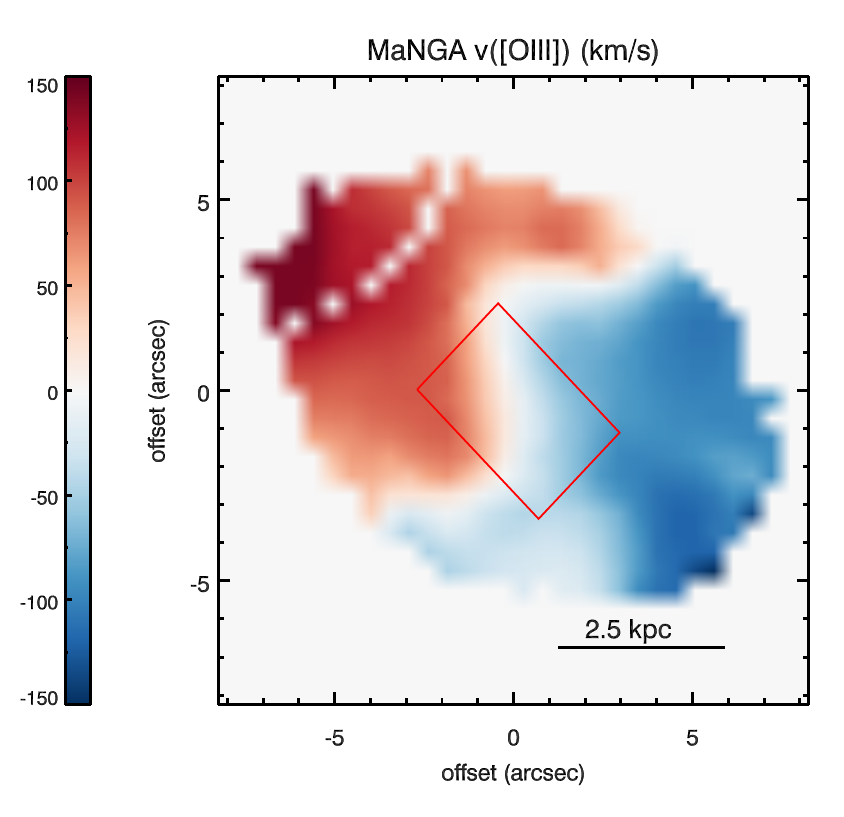}
\includegraphics[scale = 0.28, trim = 1.4cm 3.1cm 6cm 11cm, clip = true]{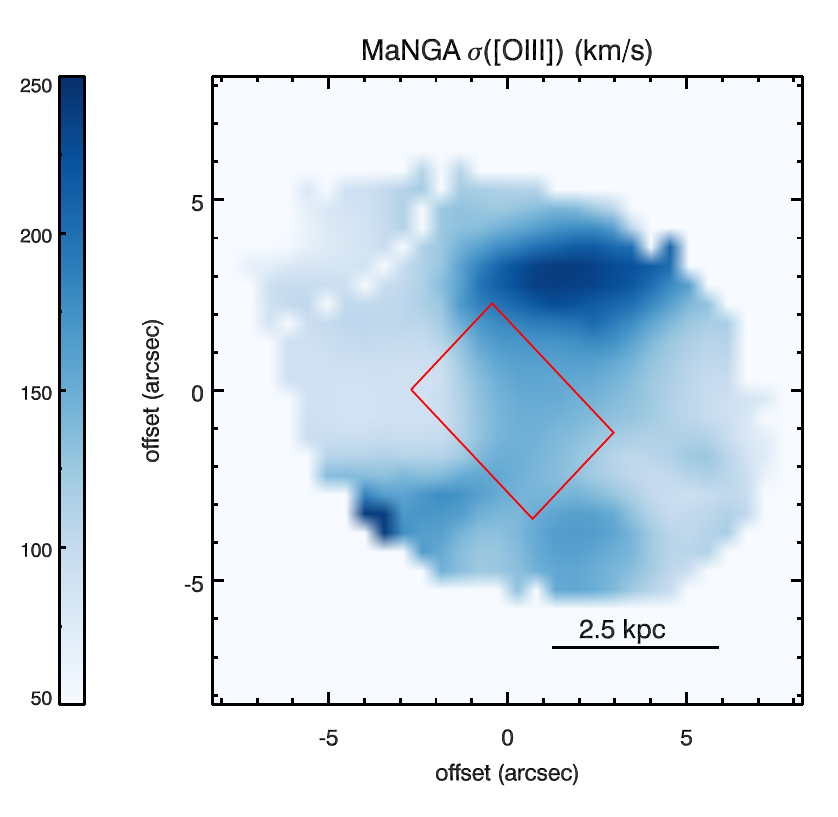}
\includegraphics[scale = 0.28, trim = 1.3cm 3.1cm 6cm 14cm, clip = true]{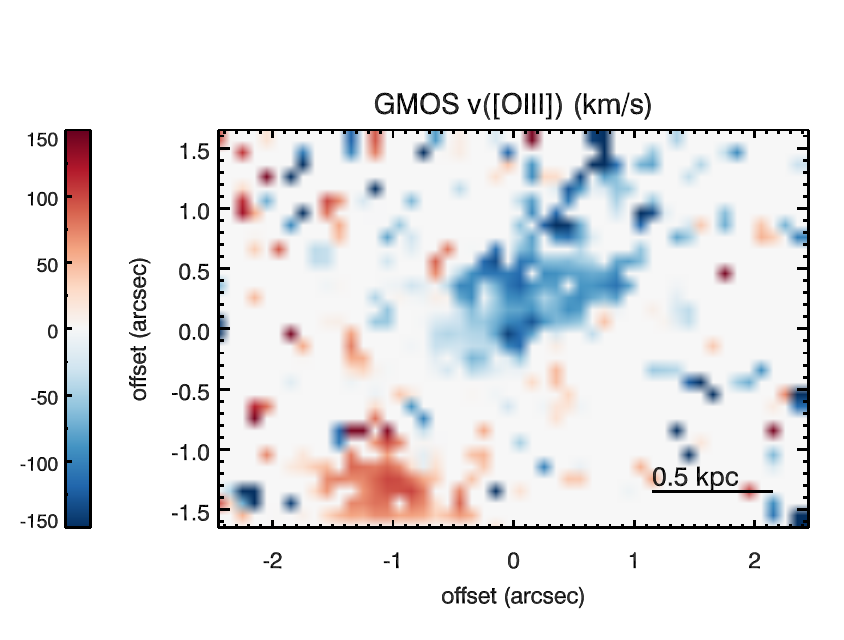}
\includegraphics[scale = 0.28, trim = 1.4cm 3.1cm 6cm 14cm, clip = true]{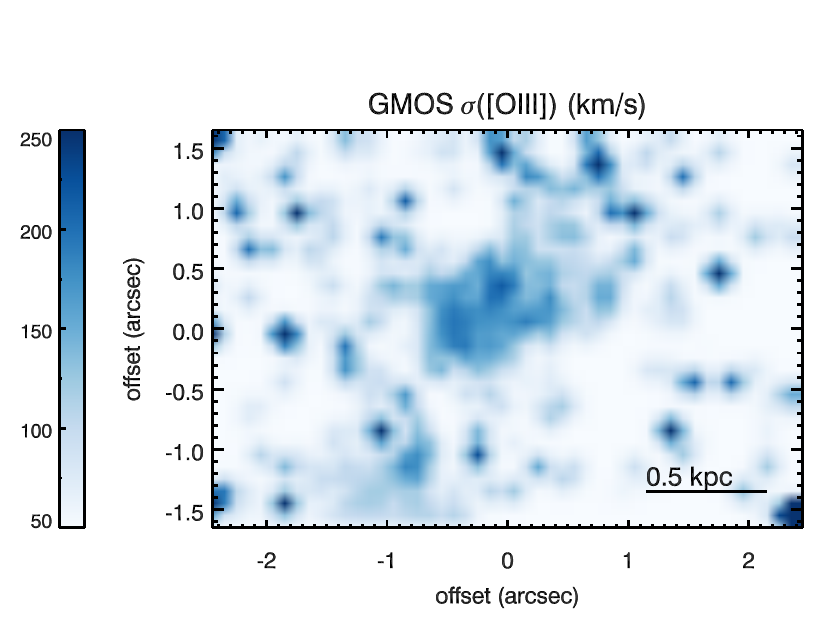}
\caption{[OIII] velocity and velocity dispersion map of the Cone Source as observed with MaNGA and GMOS. The morphology and high velocity dispersion of the cone-like structure that is apparent in both the MaNGA and GMOS [OIII] maps suggest that this is indeed a biconical outflow that has already expanded to galaxy-wide scales of $\sim 8$~kpc. We note that the GMOS $\sigma([OIII])$ corresponds to $0.4*W_{80}$([OIII]).}
\label{GMOS_cone_disp}
\end{figure}

The highly blue and redshifted sides ($\pm \sim 100$~km/s) in the residual maps additionally spatially coincide with regions of high velocity dispersion in the ionised gas, as shown in the right panels in Figure \ref{GMOS_cone_disp}. Since the outflow is nearly in the plane of the sky, the actual physical velocities are likely much higher. Although the large-scale orientation of the cone-like structure and MaNGA residuals is perpendicular to the galaxy disk (roughly North-South), the gradient of residual velocities in the centre of the galaxy is tilted with respect to the large-scale orientation ($\sim 45$ degrees East of North), indicative of a warped outflow. The GMOS H$\alpha$ residual map confirms this. Although no strong gradient is apparent there, the orientation of the highest blue-shifted residuals agrees with the central part of the MaNGA H$\alpha$ residual map.

Figure \ref{GMOS_cone_disp} shows the [OIII] ionised gas velocity field and its velocity dispersion as probed by MaNGA and GMOS. We identify kinematic distortions in the MaNGA [OIII] velocity field in the iso-velocity curves. Such distortions imply non-circular motions are present which can have several origins. Possibilities include non-circular orbits in a bar, but also outflows can cause such distortions. [OIII] is only detected in two distinct regions of the GMOS map that show blue- and redshifted velocities, respectively. The velocity dispersion, i.e. typical bulk velocity of the detected [OIII] regions in GMOS is $\sim 200$~km/s.

The dust lane clearly visible in the image of the galaxy suggests that dust extinction can play a major role in the apparent kinematics and geometry of ionised gas. We therefore investigate the effect of dust extinction through the galaxy on the morphology of outflow signatures and morphologies. The difference in dust reddening in different parts of the galaxy (i.e. high values in the plane of the disk, low values above and below the plane of the disk) could result in a seemingly biconical morphology of the high velocity dispersion [OIII] emission which is similar to the morphology of the AGN/LINER-dominated spaxels in the resolved [NII]-BPT diagram. This could be the case if the [OIII] emitting gas cloud was spherically symmetric but only illuminated along less obscured directions (i.e. above and below the plane of the disk). We therefore investigate if the morphology of the high velocity dispersion gas is truly biconical and if the lower [OIII] flux (and [OIII] velocity dispersions) along the plane of the disk can be explained by the higher reddening values in that part of the galaxy. 

We estimate the amount of reddening $A_{V}$ in the central part of the source using the H$\alpha$ and H$\beta$ emission lines from the MaNGA data. For case B recombination \citep{Osterbrock_1989} and using extinction coefficients for the Galactic extinction curve from \citet{Cardelli_1989}, we find average reddening values of $A_V \sim 3$~mag in the central 2.5~kpc and $A_V \sim 3.5$~mag in the central pixels, while in the outer parts of the disk $A_{V} \sim 1.6$~mag. Reddening values above and below the plane of the disk range between $A_{V} \sim 0.3-1.2$~mag. If the [OIII] emitting gas had a spherically symmetric morphology, then the intrinsic (extinction corrected) [OIII] flux in the plane of the disk would be similar to the intrinsic [OIII] flux below and above the plane of the stellar disk. Therefore, the expected ratio between observed [OIII] fluxes would fulfill the following relation:
\begin{equation}
A(\rm{[OIII]})_{\rm{non\_disk}} - A(\rm{[OIII]})_{\rm{disk}} = -2.5\times\log\frac{F_{\rm{[OIII], non\_disk}}}{F_{\rm{[OIII], disk}}}
\end{equation}
with $A(\rm{[OIII]})_{\rm{non\_disk}}$ and $A(\rm{[OIII]})_{\rm{disk}}$ being the reddening values for [OIII] above and below the plane of the disk and in the plane of the disk, respectively. From the A(H$\alpha$) values in those regions and using the extinction curve from \citet{Cardelli_1989}, we find that $A(\rm{[OIII]})_{\rm{non\_disk}} \sim 0.9$~mag and $A(\rm{[OIII]})_{\rm{disk}} \sim 1.8$~mag. The expected flux ratio between the observed [OIII] fluxes in those regions, $F_{\rm{[OIII], non\_disk}}$ and $F_{\rm{[OIII], disk}}$, respectively, would therefore be $\sim 2.3$, whereas we observe flux ratios that are about 30\% higher than the expected flux ratios. This means that the [OIII] flux in the disk regions is much lower than would be expected `just' due to reddening. Lower [OIII] flux values (and lower [OIII] velocity dispersions) are therefore intrinsic and not only caused by higher extinction through the galactic disk confirming its cone-like morphology.

The morphology and velocity dispersion of the high velocity dispersion gas that is apparent in both the MaNGA and GMOS H$\alpha$ and [OIII] maps (see Figure \ref{GMOS_cone_vel} and \ref{GMOS_cone_disp}) suggests that we indeed observe a biconical outflow that has already expanded to galaxy-wide scales of $\sim 5$~kpc. This structure, however, seems to be warped in the centre of the galaxy as revealed by the MaNGA H$\alpha$ residual map and the GMOS [OIII] velocity map. Similar to the analysis described in Section 4.1, we fit a biconical outflow superimposed on gas disk rotation to the MaNGA H$\alpha$ velocity field. We find a maximum extent of the two cones combined to be $\sim 4$~kpc with $v_{\rm{max}}$ of $\sim 100$~km/s and $r_t = 950$~pc. The resulting $\dot{M}_{out}$ is $14$~M$_{\sun}$yr$^-1$ and the total kinetic power is estimated to be $\dot{E}_{\rm{kin}} = 0.5\times10^{41}$~erg s$^{-1}$. We further discuss these estimates and the connection between the small-scale GMOS observations with the larger-scale MaNGA observations in Section 4.2.

\vspace{2cm}

\section{Nature of the outflows}

We have shown that both sources presented in this paper host regions of high velocity dispersion that are distinct from the gas disk velocity field. They show a biconical morphology and kinematics are indicative of moderately fast outflows with speeds of $\sim 100-200$~km/s. In the case of the Blob Source these structures were not seen in the MaNGA maps before and are only revealed at the higher resolution of GMOS observations. The remaining question now regards the powering and nature of these outflows. 

In Section 2.1, we have summarised the use and power of BPT diagrams. The [NII]-BPT which we used for initial target selection allows to characterise between star-formation, AGN+LINER, or composite dominated emission line regions. The BPT-[SII] further allows the distinction between AGN \textit{or} LINER in addition to star-formation dominated emission line regions. LINER emission was long thought to be associated with weak AGN, but it has been shown that other ionization sources, such as ionization through hot evolved stars, can also be responsible for observed line ratios \citep{Eracleous_2010, Singh_2013, Belfiore_2016}. These discussions were in part triggered through the fact that LINER-like emission is now increasingly often observed on much larger scales (several kpc) than expected from a weak central AGN. Therefore careful analysis is required when LINER-like emission line regions are observed and such sources cannot necessarily be associated with weak AGN. In addition to other criteria, we therefore also invoke the [SII]-BPT in our discussion about the nature and powering of the observed outflows in the Cone and Blob Source. 

\begin{figure}
\centering
\includegraphics[scale = 0.43, trim = 3cm 0cm 3cm 0cm, clip =true]{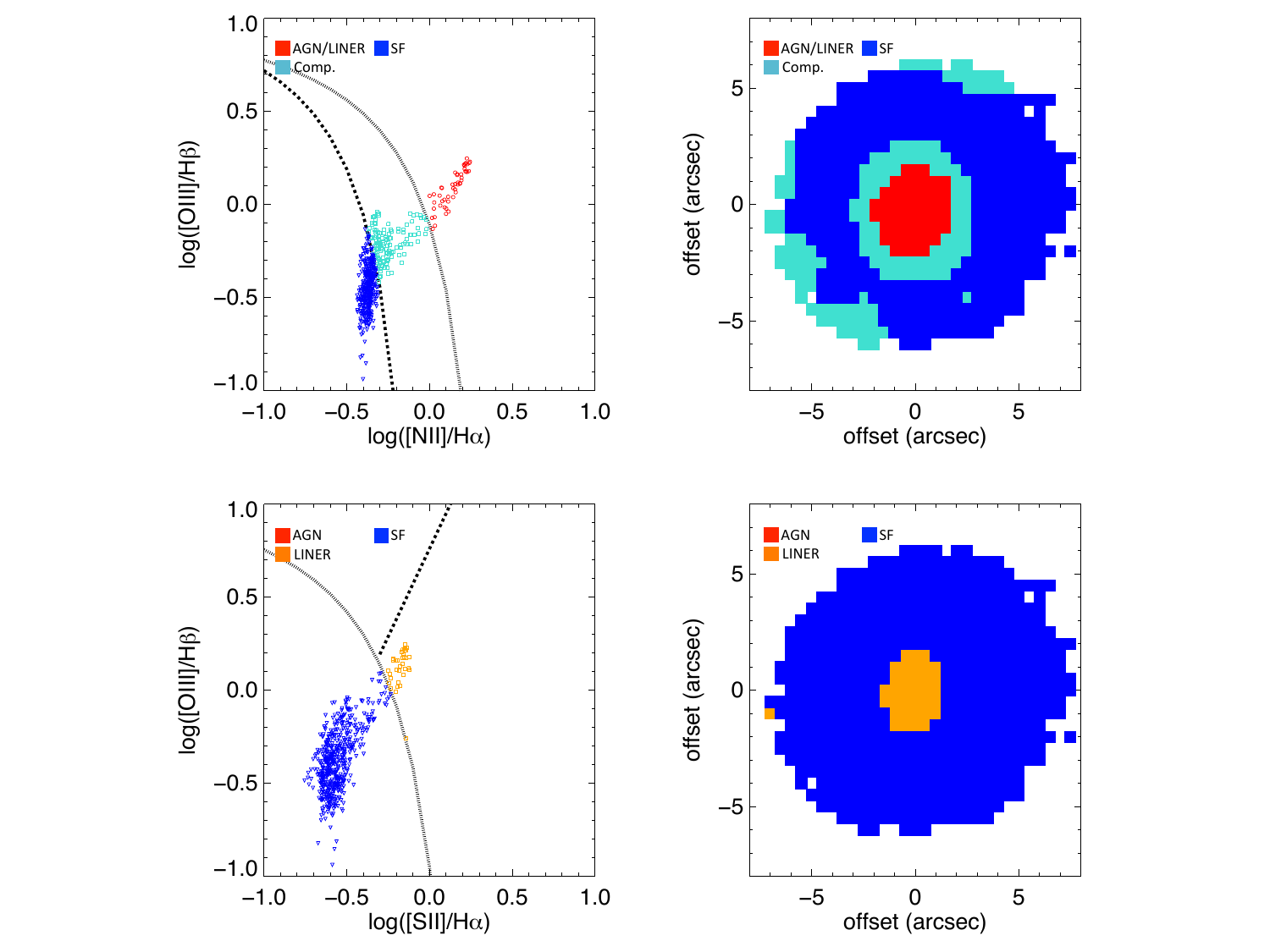}
\caption{[NII] (upper row) and [SII] (lower low) BPT diagrams of the Blob source based on the MaNGA data. We show both the position of the line ratios in the $\log$([OIII]/H$\beta$) vs. $\log$([NII]/H$\alpha$) and $\log$([OIII]/H$\beta$) vs. $\log$([SII]6584/H$\alpha$) space (left panels) and the spatially resolved BPT diagrams, where we colour code the spaxels according to their classification \citep[right panels, the dividing lines are from][]{Kewley_2006}. While the resolved [NII] BPT diagram shows that line ratios in the centre of the Blob Source show elevated line ratios which could be due to AGN photo-ionization or LINER-like ionization, the resolved [SII] BPT diagram reveals that the line ratios in the centre are LINER-like. Based on a number of arguments (see text) we show that the line ratios can best be explained by shocks created by as small-scale AGN-driven jets or outflows propagating through the gaseous stellar disk.}
\label{Blob_BPT}
\end{figure}

\subsection{Blob Source}

The situation in the Blob Source is ambiguous. According to the resolved MaNGA [SII]-BPT diagram (Figure \ref{Blob_BPT}), the central 2 arcsec region of this source shows LINER-like line ratios. \citet{Belfiore_2016} recently presented a census of LI(N)ER emission of all galaxies observed in MaNGA (as of April 2015). They chose the term LIER emission (leaving out the N for nuclear) since many galaxies show LIER dominated emission line regions on large, galaxy-wide scales, which they refer to as `extended LIER', eLIER, galaxies. A similar fraction of about 30\% of galaxies in MaNGA shows LIER emission at small galactocentric radii which \citet{Belfiore_2016} classify as `central LIER' (cLIER) galaxies. The Blob Source in this work would thus be classified as a cLIER based on the morphology and extent of the LIER-like emission in the MaNGA maps. \citet{Belfiore_2016} argue that post asymptotic giant branch (pAGB) stars can produce the required hard ionizing spectrum to power the emission in both cLIER and eLIER galaxies in MaNGA \citep{Eracleous_2010, Binette_1994}. Without the additional GMOS data that reveal that the central part of the galaxy is actually hosting a small-scale outflow, this source would have probably been classified as one of such cLIER galaxies in MaNGA. Our additional data from GMOS and kinematic analysis shows, however, that the pAGB scenario cannot be easily adapted in this case. 

The conclusion that the ionization properties in cLIER and eLIER galaxies in MaNGA are consistent with being due to ionization through pAGB stars \citep{Belfiore_2016} is mostly based on the observation of relatively low H$\alpha$ equivalent widths, $0.5 <$EW(H$\alpha$)$< 3 \AA$, in LIER emission regions which can be well reproduced by stellar evolution models \citep{Binette_1994, Cid-Fernandes_2011}. In AGN or star forming regions much higher EW(H$\alpha$) ($> 3\AA$) would be expected. The mean H$\alpha$ equivalent width in the central region with radius $\sim 3.4$~kpc of the Blob Source as measured from the MaNGA data is EW(H$\alpha$)~$= 6.6\pm0.5\AA$, much higher than would be expected for ionization through pAGB stars \citep{Cid-Fernandes_2011}. \citet{Belfiore_2016} suggest that such high EW(H$\alpha$) in LIER regions, which in their sample often correspond to localised features (often with a bisymmetric morphology) can be due to the relative geometry of ionizing stellar regions and gas cloud absorbers. The GMOS data show that the central feature in the Blob Source is actually a bicone but in MaNGA the morphology of the feature in the Blob Source is smooth and round. This makes a direct comparison with the observations by \citet{Belfiore_2016} challenging, but without the additional data from GMOS, we would conclude that a geometry-related explanation is unlikely in this case. The biconical nature of this feature on much smaller scales than resolvable with MaNGA and high EW(H$\alpha$) values lead us to conclude that the most likely explanation for the LIER-like line ratios in the Blob Source are shocks. Shocks models can reproduce the observed EW(H$\alpha$) values and are often accompanied with an increase in gas velocity dispersion such as is observed in the Blob Source \citep{Ho_2014}.

What is driving these shocks? Shocks can be associated with SF through jet and bipolar outflows in the early stages of low-mass star formation and through supernova feedback \citep{Soto_2012, Bonnell_2013}. Such shocks can then accelerate particles producing synchrotron emission that can be detected at radio wavelengths. This is especially true in highly star-forming and merging galaxies such as ultra-luminous IR galaxies \citep[ULIRGSs,][]{Soto_2012, Soto_2012b}. The Blob galaxy was detected within the Faint Images of the Radio Sky at Twenty-Centimeters \citep[FIRST,][]{Becker_1995} survey at 1.4~GHz as an unresolved source with F$_{1.4\rm{GHz}} = 2.6$~mJy. If the dominant source of the shocks in the Blob galaxy was SF, then SF is also expected to explain the amount of radio emission observed in this source. We compute the expected SFR due to radio emission using the correlation derived by \citet{Bell_2003}
\begin{equation}
\log L_{\rm{radio,SF}}[\rm{erg~s^{-1}}] = 26.4687 + 1.1054 \log(L_{\rm{IR,SF}}/L_{\sun}).
\end{equation}
Using $SFR = \frac{L_{\rm{IR,SF}}}{5.8\times10^9 L_{\sun}} $ \citep{Kennicutt_1998} this relation can be expressed as
\begin{equation}
\log L_{\rm{radio,SF}}[\rm{erg~s^{-1}}] = 37.2616+1.1054 \log(\rm{SFR})
\end{equation}
The radio luminosity of the Blob galaxy of $\nu L_{\nu}\rm{[1.4GHz]} = 4.6\times10^{38}$~erg/s implies a star formation rate of 19 M$_{\sun}$/yr whereas star formation rate estimated based on the H$\alpha$ luminosity \citep{Kennicutt_1998} is only SFR $\sim 1 $M$_{\sun}$/yr. 

The H$\alpha$-based SFR can be both contaminated by contributions from an AGN (i.e. increased) or due to extinction (i.e. decreased) and therefore only serves as a rough estimate. A more robust SFR estimate that is not affected by extinction can be obtained using mid-IR continuum measurements \citep{Wu_2005}. The Blob Source was detected with the \textit{Wide-field Infrared Survey Explorer} \citep[WISE][]{Wright_2010} satellite that imaged the whole sky at mid-infrared (IR) wavelengths. We power-law extrapolate the WISE fluxes at $12$ and $22\mu$m to calculate $S_{24\mu\rm{m}}$ and follow \citet{Wu_2005} to estimate a SFR $\sim3$M$_{\sun}$/yr based on $\nu$L$_{\nu}\rm{[24\mu m]}$. In AGN, $\nu$L$_{\nu}\rm{[24\mu m]}$ can be contaminated by hot dust emission due to AGN dust heating. This measurement therefore serves as an upper limit on the actual SFR in the galaxy. These calculations show that the SFR in the Blob source is insufficient to explain the observed radio emission by almost an order of magnitude. Radio emission and shock signatures must therefore have another origin. 

In radio-loud AGN, shocks can be the result of powerful relativistic jets that inflate over-pressured cocoons providing shock ionization \citep{Croston_2007}. We therefore investigate if the radio emission in the Blob galaxy is indicative of a radio-loud AGN residing in the host. In purely star-forming systems, the radio and IR emission are tightly correlated \citep{Helou_1985, Condon_1992, Bell_2003} and therefore one can use the ratio between observed fluxes at radio and IR wavelengths to identify radio-loud AGN in which the radio emission is not dominated by SF and show radio excess. We follow \citet{Bonzini_2013} and use the parameter $q_{24\mu m} = \log(S_{24\mu\rm{m}}/S_{1.4\rm{GHz}})$ where $S_{24\mu\rm{m}}$ and $S_{1.4\rm{GHz}}$ are the observed fluxes at 24$\mu$m and 1.4~GHz and find $q \sim 0.5$. This puts the source right at the dividing line between radio-loud AGN and non-radio-loud AGN (i.e. radio-quiet AGN and star forming galaxies) \citep{Bonzini_2013}. So while we observe a slight radio excess in terms of the $q_{24\mu m}$ parameter, this observation alone does not allow us not firmly infer that a small scale AGN-driven radio jet is driving the shocks. 

Recently, \citet{Zakamska_2014} suggested that in radio-quiet AGN winds can be radiatively accelerated and drive shocks into the host galaxy as they propagate through the interstellar medium. They observe a statistically significant correlation between [OIII] line kinematics $W_{90}$ and radio luminosity suggesting a common origin. Assuming that the efficiency of converting the kinetic energy of the outflow into radio synchrotron emission in such radiatively driven outflows is similar to starburst-driven winds ($3.6\times10^{-5}$) they show that the radio emission in their sample of luminous type-2 AGN can indeed be explained in as a bi-product of AGN-driven winds. They derive efficiencies of a few percent for converting the bolometric AGN luminosity to the kinetic luminosity of the wind which agrees with previous estimates based on observations of spatially resolved winds \citep{Liu_2013b}. Applying the same wind-to-radio emission efficiency to the Blob source, we require $L_{\rm{wind}} = 1.3\times10^{43}$erg/s to reproduce the observed radio luminosity of $\nu L_{\nu}\rm{[1.4GHz]} = 4.6\times10^{38}$erg/s. This estimate is about a magnitude higher than the kinetic power of the outflow we derive in Section 3.1. But, as mentioned in Section 3.1, there are large uncertainties involved in deriving the kinetic power of an outflow just from ionised gas observations alone \citep[see also][]{Muller-Sanchez_2011, Liu_2013b, Kakkad_2016}. Because of the large uncertainties involved in both estimates, we consider them to be roughly consistent with each other. This shows that an AGN-driven outflow in the Blob Source can account both for the observed radio emission and the ionised gas kinematics.

If the Blob source indeed hosts an AGN, we can estimate the bolometric AGN luminosity from $L_{\rm{[OIII]}} = 2.5\times10^{40}$erg/s \citep{Reyes_2008}. In low- to intermediate-luminosity AGN, the [OIII] luminosity can be impacted by contributions from star formation. Using 400 galaxies at $0.01 < z < 0.07$ from the SDSS data release 7 \citep{Abazajian_2009}, \citet{Wild_2010} developed a method to estimate the star formation contributions to $L_{\rm{[OIII]}}$ based on the location of the source in the BPT diagram. Following this analysis, we estimate that the SF contribution to $L_{\rm{[OIII]}}$ is $< 15\%$. We therefore estimate that $L_{\rm{bol, AGN}} \sim 6 -7.1\times10^{43}$erg/s. This would imply efficiencies of converting $L_{\rm{bol, AGN}}$ to $L_{\rm{wind}}$ of $\sim 3-15$\%. This is a higher efficiency than derived by \citet{Zakamska_2014}. The L$_{\rm{bol, AGN}}$-to-$L_{\rm{wind}}$-efficiency in \citet{Zakamska_2014} was, however, derived for much more luminous quasars with high-velocity winds on galaxy-wide (tens of kpc) scale. In the outer parts of the galaxy, where gas densities are lower, lower efficiencies are naturally expected \citep{Dugan_2016}. Therefore, it is possible that the average efficiency in larger-scale winds is lower than in young small-scale outflows in higher density environments. If an AGN resides in the centre of the Blob galaxy, the kinetic luminosity of the wind is therefore sufficient to power the radio emission in the source which in turn can be associated with the shock signatures in the centre of the galaxy.

The dynamical timescale of the ionised gas outflow ($t_{\rm{dyn}} = r_t/v_{\rm{max}}$) is $2-3$~Myr. Typical AGN lifetimes have been estimated through various methods in the past to be on the order of a few $10^7-10^8$~yr \citep{Haehnelt_1993}. The spatial extent of the outflow can therefore be easily generated through a young AGN that has only been active for a few Myr \citep[see also][]{Sun_2014}. Furthermore, the observed radio luminosity for the Blob galaxy agrees with the radio luminosity that would be expected based on the correlation between $W_{90}$ and $\nu L_{\nu}\rm{[1.4GHz]}$ for luminous type-2 AGN \citep{Zakamska_2014}. This suggests that the kinematic signatures in this galaxy might be of the same origin as seen in more luminous AGN and are just the extension to younger and smaller outflows. 

The GMOS observations reveal that the morphology of the inner region of the Blob source has a biconical structure with a well-defined velocity gradient indicative of an outflow (Figure \ref{GMOS_blob}). Based on the above derived arguments we conclude that the Blob source hosts an AGN. The observed ionised gas and radio emission can then be explained either by small-scale (1-2 kpc) jets that have not yet broken out of the galaxy or by radiatively-driven winds which produce shocks in the host galaxy as they propagate through the interstellar medium. We are likely observing the shocks created by the expanding hot bubble like it is seen in recent simulations of jets/outflows propagating through gaseous stellar disks \citep{Dugan_2016}. This also means, however, that the signatures in weak AGN with small-scale outflows can appear to be LINER-like and are not obviously visible as AGN photo-ionised regions. Only with combining MaNGA and GMOS were we able to detect the outflow signatures in the centre of the source. 

\subsection{Cone Source}

The Cone Source has been classified as bona-fide AGN using different selection criteria including X-ray observations \citep{Ajello_2012}. The resolved [SII]-BPT maps based on the MaNGA data show that the emission-line ratios in the cone-like structure are consistent with being ionised by an AGN. But while it is clear that the ionization itself is not due to star formation but due to an AGN, star formation can drive outflow activity and could be responsible for the observed velocity dispersion. \citet{Heckman_2015} shows that gas velocities in starburst-driven outflows correlate with the overall star formation rate (SFR) in the galaxy. We therefore estimate the star formation in the Cone Source's host galaxy using the H$\alpha$ emission line \citep{Kennicutt_1998}. The H$\alpha$ flux can be contaminated by the AGN which would increase the estimated SFR and by extinction which would lower the estimated SFR, so that a H$\alpha$-based SFR calculation can only be regarded as a rough estimate. Based on the H$\alpha$ luminosity, we find a SFR $\sim 1$M$_{\sun}$/year for the Cone Source's host galaxy. We also cross-match the Cone Source's position with the FIRST radio catalog and find a match with an unresolved source with F$_{1.4\rm{GHz}} = 3.4$~mJy which implies a radio luminosity of $\nu L_{\nu}\rm{[1.4GHz]} = 7.9\times10^{37}$~erg/s. Since the Cone Source does host a bona-fide AGN, this radio luminosity is likely impacted by contributions of the AGN, so a radio-based SFR estimate serves as an upper limit to the true SFR. Following the same arguments as in Section 4.1, we estimate that SFR$< 4$M$_{\sun}$/year.

At SFR that low (or even lower, since that is an upper limit), a starburst driven outflow is not expected to exceed velocities of $\sim 50$~km/s \citep{Martin_2005, Heckman_2015} but the outflow velocity in the biconical region reaches $100-200$~km/s (see [OIII] velocity dispersion maps in Figure \ref{GMOS_cone_disp} and outflow modelling described in Section 3.2) which by far exceeds the expected value due to star formation and implies that it is the AGN that is driving the observed outflow. 

To assess if the AGN is powerful enough to drive the observed outflow, we again estimate the bolometric luminosity of the AGN from $L_{\rm{[OIII]}} = 4\times10^{40}$erg/s and find $L_{\rm{bol, AGN}} \sim 9\times10^{43}$erg/s \citep{Reyes_2008}. In Section 3.2 we have shown that the kinetic power of the outflow is $\dot{E}_{\rm{kin}} = 0.5 \times 10^{41}$erg~s$^{-1}$. This implies an efficiency ($\dot{E}_{\rm{kin}}/L_{\rm{bol, AGN}}$) of $< 0.1\%$. The dynamical timescale of the outflow $t_{\rm{dyn}}$ is in the range of $7-10$~Myr, consistent with the typical active phase of an AGN. These arguments show that the AGN is more than powerful and long-lasting enough to produce the ionised gas outflows observed. 

Interesting to note is the already mentioned directional change of the outflow that occurs at a distance of $\sim 2$~\arcsec\ from the centre of the Cone Source. This directional change also becomes apparent in the warped iso-velocity contours of $\sim 0$~km/s in the MaNGA [OIII] velocity field (Figure \ref{GMOS_cone_disp}). The fact that we observe [OIII] in only two distinct regions in the GMOS field of view that are aligned almost perpendicular to the large-scale outflow direction further confirms this observation. Such a directional change in kinematics could, however, also be related to the presence of further kinematically distinct components in the host galaxy, such as a central bar. But the high [OIII] velocity dispersions ($\sim 200$~km/s) in the centre of the galaxy as probed with GMOS are unlikely to be explained by stellar processes. This suggests that the high velocity gas in the centre of the galaxy and the outer parts of the galaxy are indeed kinematically and physically associated. The directional change of the outflow is possibly connected to the `path of least resistance' through the host galaxy that outflows have been shown to prefer \citep{Faucher_2012}. Another possibility is that above and below the plane of the galaxy, where extinction is low, the outflow is propagating rapidly and is ionised out to far distances. In the disk, where extinction is higher, the outflow is slowed down and due to higher extinction values, we predominantly detect the blueshifted part of the outflow (Figure \ref{GMOS_cone_disp}), leading to this seemingly directional change of the outflow. The observed line ratios further show that the dominant ionization source is an AGN.

This analysis strongly suggests that the cone-like structure is indeed powered and driven by the AGN. Its biconical nature and velocity field reveal that it is a moderately fast outflow expanding parallel to the stellar disk on small scales and perpendicular to the stellar disk on large scales.

\section{Additional impact of GMOS observations}

In the previous subsections we have shown that we observe signatures of outflows that are driven by a central AGN in both the Blob and the Cone Source. While in the Cone Source, the biconical outflow signatures were already apparent in the MaNGA observations and were observed on galactic-wide scales, the outflow signatures in the Blob Source could only be resolved with the higher-resolution GMOS observations. We now investigate the how the differences in outflow signatures (small-scale vs. large-scale) may relate to their host galaxy parameters.

Although $L_{\rm{[OIII]}}$ can be affected by extinction, most likely from dust in the narrow-line region, [OIII] luminosities are a good indicator of total bolometric AGN luminosity \citep{Reyes_2008, LaMassa_2010}. These relations have been established using the single fibre SDSS spectra and we therefore measure [OIII] luminosities for the Cone and Blob source using the single fibre spectra for this part of the analysis. Interestingly we find a slightly smaller [OIII] luminosity in the Blob Source (log(L[OIII]/(erg s$^{-1}$)) $\sim 40.4$), in which we suspect the smaller/younger outflow, than in the Cone Source (log(L[OIII]/(erg s$^{-1}$)) $\sim 40.6$). The total bolometric luminosities of the Blob and Cone Source are therefore $\sim 10^{43-44}$~erg/s. Recent work \citep{Veilleux_2013, Zakamska_2014, Wylezalek_2016b} suggests that there is a tight correlation between bolometric AGN luminosity and the signatures of feedback such as outflow velocities quantified by emission line width measurements. Such a correlation then implies that there is a minimum AGN luminosity above which an AGN can become powerful enough to launch winds that will affect the galaxy. This threshold luminosity has been estimated to be $L_{\rm{bol}} \sim 10^{45}$~erg/s. In AGN with $10^{43}<L_{\rm bol}< 10^{45}$ erg/s key AGN feeding and feedback processes are expected to occur within the inner kpc \citep{Storchi_2010,Barbosa_2014, Lena_2015}. The two sources in this work are around or slightly below the expected threshold luminosity and interestingly we observe the more extended and more developed outflow signatures in the higher luminosity object, which is in line with expectations from these recent works. Additionally, the Cone host galaxy stellar mass is almost an order of magnitude smaller than the Blob Source's stellar mass ($1\times10^{10}$M$_{\sun}$ for the Cone Source and $7.3\times10^{10}$M$_{\sun}$ for the Blob Source). Stellar masses have been estimated from $K-$correction fits to the optical photometric data from SDSS and are reported in the NSA Sloan Atlas catalogue. This potentially points to the fact that the outflow in the Cone Source had to overcome a smaller gravitational potential facilitating expansion to larger galactic scales in comparison with the outflow in the Blob Source. In an upcoming Gemini GMOS program which will observe a larger sample of MaNGA-AGN with a similar observational setup, we will be investigating the correlation between outflow signatures, AGN luminosity and galactic potential on a larger statistical basis. Outflows and feedback from low and intermediate luminosity AGN can potentially have significant impact on their environment and contribute to AGN/host-galaxy self-regulation \citep{Crenshaw_2012}.

\section{Conclusions}

In this paper, we combine IFU observations obtained within the SDSS-IV MaNGA survey with higher resolution, small field-of-view IFU observations obtained with the Gemini/GMOS instrument of two low-redshift AGN with luminosities about an order of magnitude below the traditional quasar cut-off. While the MaNGA observations show the surface brightness distribution and kinematics of gas and stars on galactic wide scales, the GMOS observations zoom into the centre and resolve the inner few kpc of the galaxies. 

Both sources have been selected from the MaNGA sample based on their very different ionised gas morphologies (cone-like vs. blob-like) and kinematics and on their high fraction of highly ionised regions as probed by resolved MaNGA [NII]-BPT diagrams. The goal of this work is to map circumnuclear gas flows on scales from $\la 250$~pc to $\sim 2$~kpc in AGN whose host galaxies are mapped by MaNGA and where we expect to detect the long-sought transition from circumnuclear to galaxy-wide AGN feedback.

Revealing the nature of the Blob Source is challenging due to the morphology and ambiguous line ratios in the central region of the source. Combining IFU observations with different resolutions and mapping different spatial scales, we find the following: 

\begin{itemize}
\item The Blob Source shows a circular region with radius $\sim 2.7$~kpc of AGN/LINER-dominated spaxels in the resolved MaNGA [NII] BPT diagram that is spatially coincident with increased [OIII] and H$\alpha$ velocity dispersion.
\item Modelling the GMOS H$\alpha$ velocity field in the inner 5~kpc of the Blob Source reveals a biconical structure in the residual map. This biconical structure is also spatially coincident with a cone-like structure of increased H$\alpha$ velocity dispersion and [NII]/H$\alpha$ line ratios. 
\item The resolved MaNGA [SII] diagram shows that the emission line ratios in the central part of the Blob Source are LI(N)ER-like. Due to high EW(H$\alpha$) values of $\sim 6.5\AA$, we suggest that the most likely explanation for the LIER-like line ratios in the Blob Source are shocks. 
\item Based on a multi-wavelength analysis involving estimates of (kinetic) energies of the shock and SF- and AGN-driven winds, we conclude that the structure represents a small or stalled outflow or jet powered by an AGN which is driving the shocks. This conclusion is backed by the GMOS observations which resolve the biconical nature of that region with a well-defined velocity gradient in the outflow.
\end{itemize} 

\noindent We find the following in the Cone Source:

\begin{itemize}
\item The MaNGA maps of the Cone Source show a biconical region of high [OIII] velocity dispersion that is spatially coincident with AGN-dominated spaxels in both the MaNGA resolved [NII] and [SII] BPT diagrams.
\item Modelling of MaNGA H$\alpha$ velocity maps reveals an additional biconical kinematic component that is spatially coincident with the high [OIII] velocity dispersion region. However, the decoupled kinematic component is warped and changes direction at a distance of $\sim 1$~kpc from the centre. Such a change in direction could have multiple origins, including the presence of a stellar bar in the centre of the galaxy.
\item The GMOS observations of the inner few kpc of the Cone Source reveal high velocity dispersion [OIII] emission that is detected in two regions in the GMOS field of view. The alignment is in agreement with the MaNGA observations and its high velocity dispersion makes a bar scenario unlikely. 
\item We suggest that this cone-like structure that shows a clear velocity gradient in the residual maps and emission line ratios consistent with being excited by an AGN represents a moderately fast outflow that has already propagated to galaxy-wide scales with a size of several kpc. This conclusion is in agreement with calculations based on the kinematics and energetics in the outflow and of the AGN.
\end{itemize}

Some authors have recently argued that the kinematics of the narrow-line region of some AGN can be largely due to a combination of rotation and in situ acceleration of material originating in the host disk as an alternative for material outflowing from the nuclear regions. The low/intermediate luminosities ($10^{43}-10^{44}$~erg~s$^{-1}$) of the AGN in this work, combined with the large distances at which we detect outflowing components that are kinematically decoupled from the stellar gas disk ($2-4$~kpc) are hard to reconcile with models in which gas in the galactic disk is accelerated in situ at these distances \citep{Fischer_2016}. The conical geometry of our outflow constrains the acceleration size to be $<400-700$~pc, the typical resolution of our observations. Higher resolution observations would be required to trace the outflow to its footprints within the nucleus of the galaxy and to determine the size of the acceleration region.

We therefore conclude that we have detected AGN-driven outflows in both targets. In the Cone Source, the higher-$L_{\rm{bol}}$, lower stellar mass object, we are indeed tracing the origin and coupling of the wind to the large-scale biconical outflow and resolve the outflow direction with GMOS. But the MaNGA maps reveal how far the outflow reaches and its large-scale propagation direction. In the Blob Source, the lower-$L_{\rm{bol}}$, higher stellar mass object, the data enabled us to discover a young or stalled biconical outflow where none was obvious at the MaNGA resolution. But the spectral range and depth of the MaNGA spectra pointed us towards further investigating the nature of the blob-like distinct structure in the centre of the galaxy. With MaNGA alone, this source would have been classified as a regular (c)LIER galaxy but with the additional kinematic information from GMOS that led to a careful multi-wavelength analysis were we able to investigate the nature of this source.

Constraining the kinetic energy and momentum of such AGN-driven outflows is still a challenging task with the available observations. \citet{Stern_2016} summarise observational constraints for such measurements for various observational techniques. They find that the ratio of outflow momentum $\dot{p}$ to the available momentum of AGN photons $L_{\rm bol}/c$ is $\gg 1$ for the narrow-line gas observations on large scales. They argue that this observation can be reasonably explained by an energy-conserving flow which follows after the wind builds up circum-nuclear pressure as it runs into the interstellar medium of the galaxy for the first time. Our outflow kinetic energy estimates presented in Sections 3.1 and 3.2 yield $\dot{p}/(L_{\rm bol}/c)$ values $\ga 100$, in general agreement with the other narrow-line-based estimates compiled by \citet{Stern_2016}. Unfortunately, all these estimates are affected by strong biases and assumptions that we mentioned in Sections 3.1 and 4.1; in particular, the electron density is very poorly known. Detections of the other phases of the same outflow (e.g., the associated neutral or molecular gas) would be an ideal avenue for testing these widely-used estimates using a different technique.

These observations show that combining large-scale IFU data with higher resolution, small scale IFU maps are necessary when exploring how wind launching and propagation are related to AGN luminosity and galaxy potential. Small young or stalled outflows might have been missed in previous observations due to resolution-limited observations.  In a forthcoming paper, we will explore the relation between outflow size, ionization mechanisms, AGN luminosity and galaxy potential with a larger set of AGN that will have been observed by both MaNGA and GMOS. Outflows and feedback from low-luminosity AGN might potentially contribute significantly to feedback processes in the galaxy. Finding hidden and small-scale outflow is therefore crucial for further understanding AGN/host galaxy self-regulation.

\section*{Acknowledgements}
D.W. acknowledges support by the Akbari-Mack Postdoctoral Fellowship and the JHU Provost's Postdoctoral Diversity Fellowship. R.A.R. acknowledges support from FAPERGS (project N0. 2366-2551/14-0) and CNPq (project N0. 470090/2013-8 and 302683/2013. G. L. is supported by the National Thousand Young Talents Program of China, and acknowledges the grant from the National Natural Science Foundation of China (No. 11673020 and No. 11421303) and the Ministry of Science and Technology of China (National Key Program for Science and Technology Research and Development, No. 2016YFA0400700).

Funding for the Sloan Digital Sky Survey IV has been provided by
the Alfred P. Sloan Foundation, the U.S. Department of Energy Office of
Science, and the Participating Institutions. SDSS-IV acknowledges
support and resources from the Center for High-Performance Computing at
the University of Utah. The SDSS web site is www.sdss.org.

SDSS-IV is managed by the Astrophysical Research Consortium for the 
Participating Institutions of the SDSS Collaboration including the 
Brazilian Participation Group, the Carnegie Institution for Science, 
Carnegie Mellon University, the Chilean Participation Group, the French Participation Group, Harvard-Smithsonian Center for Astrophysics, 
Instituto de Astrof\'isica de Canarias, The Johns Hopkins University, 
Kavli Institute for the Physics and Mathematics of the Universe (IPMU) / 
University of Tokyo, Lawrence Berkeley National Laboratory, 
Leibniz Institut f\"ur Astrophysik Potsdam (AIP),  
Max-Planck-Institut f\"ur Astronomie (MPIA Heidelberg), 
Max-Planck-Institut f\"ur Astrophysik (MPA Garching), 
Max-Planck-Institut f\"ur Extraterrestrische Physik (MPE), 
National Astronomical Observatory of China, New Mexico State University, 
New York University, University of Notre Dame, 
Observat\'ario Nacional / MCTI, The Ohio State University, 
Pennsylvania State University, Shanghai Astronomical Observatory, 
United Kingdom Participation Group,
Universidad Nacional Aut\'onoma de M\'exico, University of Arizona, 
University of Colorado Boulder, University of Oxford, University of Portsmouth, 
University of Utah, University of Virginia, University of Washington, University of Wisconsin, 
Vanderbilt University, and Yale University.




\bibliographystyle{mnras}
\bibliography{master_bib} 

\begin{thebibliography}{}
\makeatletter
\relax
\def\mn@urlcharsother{\let\do\@makeother \do\$\do\&\do\#\do\^\do\_\do\%\do\~}
\def\mn@doi{\begingroup\mn@urlcharsother \@ifnextchar [ {\mn@doi@}
  {\mn@doi@[]}}
\def\mn@doi@[#1]#2{\def\@tempa{#1}\ifx\@tempa\@empty \href
  {http://dx.doi.org/#2} {doi:#2}\else \href {http://dx.doi.org/#2} {#1}\fi
  \endgroup}
\def\mn@eprint#1#2{\mn@eprint@#1:#2::\@nil}
\def\mn@eprint@arXiv#1{\href {http://arxiv.org/abs/#1} {{\tt arXiv:#1}}}
\def\mn@eprint@dblp#1{\href {http://dblp.uni-trier.de/rec/bibtex/#1.xml}
  {dblp:#1}}
\def\mn@eprint@#1:#2:#3:#4\@nil{\def\@tempa {#1}\def\@tempb {#2}\def\@tempc
  {#3}\ifx \@tempc \@empty \let \@tempc \@tempb \let \@tempb \@tempa \fi \ifx
  \@tempb \@empty \def\@tempb {arXiv}\fi \@ifundefined
  {mn@eprint@\@tempb}{\@tempb:\@tempc}{\expandafter \expandafter \csname
  mn@eprint@\@tempb\endcsname \expandafter{\@tempc}}}

\bibitem[\protect\citeauthoryear{{Abazajian} et~al.,}{{Abazajian}
  et~al.}{2009}]{Abazajian_2009}
{Abazajian} K.~N.,  et~al., 2009, \mn@doi [\apjs]
  {10.1088/0067-0049/182/2/543}, \href
  {http://adsabs.harvard.edu/abs/2009ApJS..182..543A} {182, 543}

\bibitem[\protect\citeauthoryear{{Ajello}, {Alexander}, {Greiner}, {Madejski},
  {Gehrels}  \& {Burlon}}{{Ajello} et~al.}{2012}]{Ajello_2012}
{Ajello} M.,  {Alexander} D.~M.,  {Greiner} J.,  {Madejski} G.~M.,  {Gehrels}
  N.,   {Burlon} D.,  2012, \mn@doi [\apj] {10.1088/0004-637X/749/1/21}, \href
  {http://adsabs.harvard.edu/abs/2012ApJ...749...21A} {749, 21}

\bibitem[\protect\citeauthoryear{{Baldwin}, {Phillips}  \&
  {Terlevich}}{{Baldwin} et~al.}{1981}]{Baldwin_1981}
{Baldwin} J.~A.,  {Phillips} M.~M.,   {Terlevich} R.,  1981, \mn@doi [\pasp]
  {10.1086/130766}, \href {http://adsabs.harvard.edu/abs/1981PASP...93....5B}
  {93, 5}

\bibitem[\protect\citeauthoryear{{Barbosa}, {Storchi-Bergmann}, {Cid
  Fernandes}, {Winge}  \& {Schmitt}}{{Barbosa} et~al.}{2006}]{Barbosa_2006}
{Barbosa} F.~K.~B.,  {Storchi-Bergmann} T.,  {Cid Fernandes} R.,  {Winge} C.,
  {Schmitt} H.,  2006, \mn@doi [\mnras] {10.1111/j.1365-2966.2006.10690.x},
  \href {http://adsabs.harvard.edu/abs/2006MNRAS.371..170B} {371, 170}

\bibitem[\protect\citeauthoryear{{Barbosa}, {Storchi-Bergmann}, {Cid
  Fernandes}, {Winge}  \& {Schmitt}}{{Barbosa} et~al.}{2009}]{Barbosa_2009}
{Barbosa} F.~K.~B.,  {Storchi-Bergmann} T.,  {Cid Fernandes} R.,  {Winge} C.,
  {Schmitt} H.,  2009, \mn@doi [\mnras] {10.1111/j.1365-2966.2009.14485.x},
  \href {http://adsabs.harvard.edu/abs/2009MNRAS.396....2B} {396, 2}

\bibitem[\protect\citeauthoryear{{Barbosa}, {Storchi-Bergmann}, {McGregor},
  {Vale}  \& {Rogemar Riffel}}{{Barbosa} et~al.}{2014}]{Barbosa_2014}
{Barbosa} F.~K.~B.,  {Storchi-Bergmann} T.,  {McGregor} P.,  {Vale} T.~B.,
  {Rogemar Riffel} A.,  2014, \mn@doi [\mnras] {10.1093/mnras/stu1637}, \href
  {http://adsabs.harvard.edu/abs/2014MNRAS.445.2353B} {445, 2353}

\bibitem[\protect\citeauthoryear{{Becker}, {White}  \& {Helfand}}{{Becker}
  et~al.}{1995}]{Becker_1995}
{Becker} R.~H.,  {White} R.~L.,   {Helfand} D.~J.,  1995, \mn@doi [\apj]
  {10.1086/176166}, \href {http://adsabs.harvard.edu/abs/1995ApJ...450..559B}
  {450, 559}

\bibitem[\protect\citeauthoryear{{Belfiore} et~al.,}{{Belfiore}
  et~al.}{2016}]{Belfiore_2016}
{Belfiore} F.,  et~al., 2016, \mn@doi [\mnras] {10.1093/mnras/stw1234}, \href
  {http://adsabs.harvard.edu/abs/2016MNRAS.461.3111B} {461, 3111}

\bibitem[\protect\citeauthoryear{{Bell}}{{Bell}}{2003}]{Bell_2003}
{Bell} E.~F.,  2003, \mn@doi [\apj] {10.1086/367829}, \href
  {http://adsabs.harvard.edu/abs/2003ApJ...586..794B} {586, 794}

\bibitem[\protect\citeauthoryear{{Bertola}, {Bettoni}, {Danziger}, {Sadler},
  {Sparke}  \& {de Zeeuw}}{{Bertola} et~al.}{1991}]{Bertola_1991}
{Bertola} F.,  {Bettoni} D.,  {Danziger} J.,  {Sadler} E.,  {Sparke} L.,   {de
  Zeeuw} T.,  1991, \mn@doi [\apj] {10.1086/170058}, \href
  {http://adsabs.harvard.edu/abs/1991ApJ...373..369B} {373, 369}

\bibitem[\protect\citeauthoryear{{Binette}, {Magris}, {Stasi{\'n}ska}  \&
  {Bruzual}}{{Binette} et~al.}{1994}]{Binette_1994}
{Binette} L.,  {Magris} C.~G.,  {Stasi{\'n}ska} G.,   {Bruzual} A.~G.,  1994,
  \aap, \href {http://adsabs.harvard.edu/abs/1994A%26A...292...13B} {292, 13}

\bibitem[\protect\citeauthoryear{{Blanton}}{{Blanton}}{2017}]{Blanton_2017}
{Blanton} M.~R.,  2017, in prep.

\bibitem[\protect\citeauthoryear{{Bonnell}, {Dobbs}  \& {Smith}}{{Bonnell}
  et~al.}{2013}]{Bonnell_2013}
{Bonnell} I.~A.,  {Dobbs} C.~L.,   {Smith} R.~J.,  2013, \mn@doi [\mnras]
  {10.1093/mnras/stt004}, \href
  {http://adsabs.harvard.edu/abs/2013MNRAS.430.1790B} {430, 1790}

\bibitem[\protect\citeauthoryear{{Bonzini}, {Padovani}, {Mainieri},
  {Kellermann}, {Miller}, {Rosati}, {Tozzi}  \& {Vattakunnel}}{{Bonzini}
  et~al.}{2013}]{Bonzini_2013}
{Bonzini} M.,  {Padovani} P.,  {Mainieri} V.,  {Kellermann} K.~I.,  {Miller}
  N.,  {Rosati} P.,  {Tozzi} P.,   {Vattakunnel} S.,  2013, \mn@doi [\mnras]
  {10.1093/mnras/stt1879}, \href
  {http://adsabs.harvard.edu/abs/2013MNRAS.436.3759B} {436, 3759}

\bibitem[\protect\citeauthoryear{{Bottema}, {van der Kruit}  \&
  {Freeman}}{{Bottema} et~al.}{1987}]{Bottema_1987}
{Bottema} R.,  {van der Kruit} P.~C.,   {Freeman} K.~C.,  1987, \aap, \href
  {http://adsabs.harvard.edu/abs/1987A%26A...178...77B} {178, 77}

\bibitem[\protect\citeauthoryear{{Brusa} et~al.,}{{Brusa}
  et~al.}{2015}]{Brusa_2015b}
{Brusa} M.,  et~al., 2015, \mn@doi [\aap] {10.1051/0004-6361/201425491}, \href
  {http://adsabs.harvard.edu/abs/2015A%26A...578A..11B} {578, A11}

\bibitem[\protect\citeauthoryear{{Bundy} et~al.,}{{Bundy}
  et~al.}{2015}]{Bundy_2015}
{Bundy} K.,  et~al., 2015, \mn@doi [\apj] {10.1088/0004-637X/798/1/7}, \href
  {http://adsabs.harvard.edu/abs/2015ApJ...798....7B} {798, 7}

\bibitem[\protect\citeauthoryear{{Cardelli}, {Clayton}  \& {Mathis}}{{Cardelli}
  et~al.}{1989}]{Cardelli_1989}
{Cardelli} J.~A.,  {Clayton} G.~C.,   {Mathis} J.~S.,  1989, \mn@doi [\apj]
  {10.1086/167900}, \href {http://adsabs.harvard.edu/abs/1989ApJ...345..245C}
  {345, 245}

\bibitem[\protect\citeauthoryear{{Cid Fernandes}, {Stasi{\'n}ska}, {Mateus}  \&
  {Vale Asari}}{{Cid Fernandes} et~al.}{2011}]{Cid-Fernandes_2011}
{Cid Fernandes} R.,  {Stasi{\'n}ska} G.,  {Mateus} A.,   {Vale Asari} N.,
  2011, \mn@doi [\mnras] {10.1111/j.1365-2966.2011.18244.x}, \href
  {http://adsabs.harvard.edu/abs/2011MNRAS.413.1687C} {413, 1687}

\bibitem[\protect\citeauthoryear{{Condon}}{{Condon}}{1992}]{Condon_1992}
{Condon} J.~J.,  1992, \mn@doi [\araa] {10.1146/annurev.aa.30.090192.003043},
  \href {http://adsabs.harvard.edu/abs/1992ARA%26A..30..575C} {30, 575}

\bibitem[\protect\citeauthoryear{{Crenshaw} \& {Kraemer}}{{Crenshaw} \&
  {Kraemer}}{2012}]{Crenshaw_2012}
{Crenshaw} D.~M.,  {Kraemer} S.~B.,  2012, \mn@doi [\apj]
  {10.1088/0004-637X/753/1/75}, \href
  {http://adsabs.harvard.edu/abs/2012ApJ...753...75C} {753, 75}

\bibitem[\protect\citeauthoryear{{Crenshaw}, {Schmitt}, {Kraemer}, {Mushotzky}
  \& {Dunn}}{{Crenshaw} et~al.}{2010}]{Crenshaw_2010}
{Crenshaw} D.~M.,  {Schmitt} H.~R.,  {Kraemer} S.~B.,  {Mushotzky} R.~F.,
  {Dunn} J.~P.,  2010, \mn@doi [\apj] {10.1088/0004-637X/708/1/419}, \href
  {http://adsabs.harvard.edu/abs/2010ApJ...708..419C} {708, 419}

\bibitem[\protect\citeauthoryear{{Croston}, {Kraft}  \& {Hardcastle}}{{Croston}
  et~al.}{2007}]{Croston_2007}
{Croston} J.~H.,  {Kraft} R.~P.,   {Hardcastle} M.~J.,  2007, \mn@doi [\apj]
  {10.1086/513500}, \href {http://adsabs.harvard.edu/abs/2007ApJ...660..191C}
  {660, 191}

\bibitem[\protect\citeauthoryear{{Drory} et~al.,}{{Drory}
  et~al.}{2015}]{Drory_2015}
{Drory} N.,  et~al., 2015, \mn@doi [\aj] {10.1088/0004-6256/149/2/77}, \href
  {http://adsabs.harvard.edu/abs/2015AJ....149...77D} {149, 77}

\bibitem[\protect\citeauthoryear{{Dugan}}{{Dugan}}{2016}]{Dugan_2016}
{Dugan} Z.,  2016, in prep.

\bibitem[\protect\citeauthoryear{{Eracleous}, {Hwang}  \& {Flohic}}{{Eracleous}
  et~al.}{2010}]{Eracleous_2010}
{Eracleous} M.,  {Hwang} J.~A.,   {Flohic} H.~M.~L.~G.,  2010, \mn@doi [\apj]
  {10.1088/0004-637X/711/2/796}, \href
  {http://adsabs.harvard.edu/abs/2010ApJ...711..796E} {711, 796}

\bibitem[\protect\citeauthoryear{{Faucher-Gigu{\`e}re} \&
  {Quataert}}{{Faucher-Gigu{\`e}re} \& {Quataert}}{2012}]{Faucher_2012}
{Faucher-Gigu{\`e}re} C.-A.,  {Quataert} E.,  2012, \mn@doi [\mnras]
  {10.1111/j.1365-2966.2012.21512.x}, \href
  {http://adsabs.harvard.edu/abs/2012MNRAS.425..605F} {425, 605}

\bibitem[\protect\citeauthoryear{{Ferrarese} \& {Ford}}{{Ferrarese} \&
  {Ford}}{2005}]{Ferrarese_2005}
{Ferrarese} L.,  {Ford} H.,  2005, \mn@doi [\ssr] {10.1007/s11214-005-3947-6},
  \href {http://adsabs.harvard.edu/abs/2005SSRv..116..523F} {116, 523}

\bibitem[\protect\citeauthoryear{{Fischer}, {Crenshaw}, {Kraemer}  \&
  {Schmitt}}{{Fischer} et~al.}{2013}]{Fischer_2013}
{Fischer} T.~C.,  {Crenshaw} D.~M.,  {Kraemer} S.~B.,   {Schmitt} H.~R.,  2013,
  \mn@doi [\apjs] {10.1088/0067-0049/209/1/1}, \href
  {http://adsabs.harvard.edu/abs/2013ApJS..209....1F} {209, 1}

\bibitem[\protect\citeauthoryear{{Fischer} et~al.,}{{Fischer}
  et~al.}{2016}]{Fischer_2016}
{Fischer} T.~C.,  et~al., 2016, preprint, \href
  {http://adsabs.harvard.edu/abs/2016arXiv160908927F} {} (\mn@eprint {arXiv}
  {1609.08927})

\bibitem[\protect\citeauthoryear{{Gunn} et~al.,}{{Gunn}
  et~al.}{2006}]{Gunn_2006}
{Gunn} J.~E.,  et~al., 2006, \mn@doi [\aj] {10.1086/500975}, \href
  {http://adsabs.harvard.edu/abs/2006AJ....131.2332G} {131, 2332}

\bibitem[\protect\citeauthoryear{{Haehnelt} \& {Rees}}{{Haehnelt} \&
  {Rees}}{1993}]{Haehnelt_1993}
{Haehnelt} M.~G.,  {Rees} M.~J.,  1993, \mn@doi [\mnras]
  {10.1093/mnras/263.1.168}, \href
  {http://adsabs.harvard.edu/abs/1993MNRAS.263..168H} {263, 168}

\bibitem[\protect\citeauthoryear{{Heckman}}{{Heckman}}{1980}]{Heckman_1980}
{Heckman} T.~M.,  1980, \aap, \href
  {http://adsabs.harvard.edu/abs/1980A%26A....87..152H} {87, 152}

\bibitem[\protect\citeauthoryear{{Heckman}, {Alexandroff}, {Borthakur},
  {Overzier}  \& {Leitherer}}{{Heckman} et~al.}{2015}]{Heckman_2015}
{Heckman} T.~M.,  {Alexandroff} R.~M.,  {Borthakur} S.,  {Overzier} R.,
  {Leitherer} C.,  2015, \mn@doi [\apj] {10.1088/0004-637X/809/2/147}, \href
  {http://adsabs.harvard.edu/abs/2015ApJ...809..147H} {809, 147}

\bibitem[\protect\citeauthoryear{{Helou}, {Soifer}  \&
  {Rowan-Robinson}}{{Helou} et~al.}{1985}]{Helou_1985}
{Helou} G.,  {Soifer} B.~T.,   {Rowan-Robinson} M.,  1985, \mn@doi [\apjl]
  {10.1086/184556}, \href {http://adsabs.harvard.edu/abs/1985ApJ...298L...7H}
  {298, L7}

\bibitem[\protect\citeauthoryear{{Ho}}{{Ho}}{2008}]{Ho_2008}
{Ho} L.~C.,  2008, \mn@doi [\araa] {10.1146/annurev.astro.45.051806.110546},
  \href {http://adsabs.harvard.edu/abs/2008ARA%26A..46..475H} {46, 475}

\bibitem[\protect\citeauthoryear{{Ho} et~al.,}{{Ho} et~al.}{2014}]{Ho_2014}
{Ho} I.-T.,  et~al., 2014, \mn@doi [\mnras] {10.1093/mnras/stu1653}, \href
  {http://adsabs.harvard.edu/abs/2014MNRAS.444.3894H} {444, 3894}

\bibitem[\protect\citeauthoryear{{Hopkins}}{{Hopkins}}{2012}]{Hopkins_2012}
{Hopkins} P.~F.,  2012, \mn@doi [\mnras] {10.1111/j.1745-3933.2011.01179.x},
  \href {http://adsabs.harvard.edu/abs/2012MNRAS.420L...8H} {420, L8}

\bibitem[\protect\citeauthoryear{{Kakkad} et~al.,}{{Kakkad}
  et~al.}{2016}]{Kakkad_2016}
{Kakkad} D.,  et~al., 2016, \mn@doi [\aap] {10.1051/0004-6361/201527968}, \href
  {http://adsabs.harvard.edu/abs/2016A%26A...592A.148K} {592, A148}

\bibitem[\protect\citeauthoryear{{Kauffmann} et~al.,}{{Kauffmann}
  et~al.}{2003}]{Kauffmann_2003}
{Kauffmann} G.,  et~al., 2003, \mn@doi [\mnras]
  {10.1111/j.1365-2966.2003.07154.x}, \href
  {http://adsabs.harvard.edu/abs/2003MNRAS.346.1055K} {346, 1055}

\bibitem[\protect\citeauthoryear{{Kennicutt}}{{Kennicutt}}{1998}]{Kennicutt_1998}
{Kennicutt} Jr. R.~C.,  1998, \mn@doi [ApJ] {10.1086/305588}, \href
  {http://adsabs.harvard.edu/abs/1998ApJ...498..541K} {498, 541}

\bibitem[\protect\citeauthoryear{{Kewley}, {Dopita}, {Sutherland}, {Heisler}
  \& {Trevena}}{{Kewley} et~al.}{2001}]{Kewley_2001}
{Kewley} L.~J.,  {Dopita} M.~A.,  {Sutherland} R.~S.,  {Heisler} C.~A.,
  {Trevena} J.,  2001, \mn@doi [\apj] {10.1086/321545}, \href
  {http://adsabs.harvard.edu/abs/2001ApJ...556..121K} {556, 121}

\bibitem[\protect\citeauthoryear{{Kewley}, {Groves}, {Kauffmann}  \&
  {Heckman}}{{Kewley} et~al.}{2006}]{Kewley_2006}
{Kewley} L.~J.,  {Groves} B.,  {Kauffmann} G.,   {Heckman} T.,  2006, \mn@doi
  [\mnras] {10.1111/j.1365-2966.2006.10859.x}, \href
  {http://adsabs.harvard.edu/abs/2006MNRAS.372..961K} {372, 961}

\bibitem[\protect\citeauthoryear{{Kormendy} \& {Ho}}{{Kormendy} \&
  {Ho}}{2013}]{Kormendy_2013}
{Kormendy} J.,  {Ho} L.~C.,  2013, \mn@doi [\araa]
  {10.1146/annurev-astro-082708-101811}, \href
  {http://adsabs.harvard.edu/abs/2013ARA%26A..51..511K} {51, 511}

\bibitem[\protect\citeauthoryear{{LaMassa}, {Heckman}, {Ptak}, {Martins},
  {Wild}  \& {Sonnentrucker}}{{LaMassa} et~al.}{2010}]{LaMassa_2010}
{LaMassa} S.~M.,  {Heckman} T.~M.,  {Ptak} A.,  {Martins} L.,  {Wild} V.,
  {Sonnentrucker} P.,  2010, \mn@doi [\apj] {10.1088/0004-637X/720/1/786},
  \href {http://adsabs.harvard.edu/abs/2010ApJ...720..786L} {720, 786}

\bibitem[\protect\citeauthoryear{{Law}}{{Law}}{2016}]{Law_2016}
{Law} David;~{Andrews} B. B.~M.,  2016, in prep.

\bibitem[\protect\citeauthoryear{{Law} et~al.,}{{Law} et~al.}{2015}]{Law_2015}
{Law} D.~R.,  et~al., 2015, \mn@doi [\aj] {10.1088/0004-6256/150/1/19}, \href
  {http://adsabs.harvard.edu/abs/2015AJ....150...19L} {150, 19}

\bibitem[\protect\citeauthoryear{{Lena} et~al.,}{{Lena}
  et~al.}{2015}]{Lena_2015}
{Lena} D.,  et~al., 2015, \mn@doi [\apj] {10.1088/0004-637X/806/1/84}, \href
  {http://adsabs.harvard.edu/abs/2015ApJ...806...84L} {806, 84}

\bibitem[\protect\citeauthoryear{{Liu}, {Zakamska}, {Greene}, {Nesvadba}  \&
  {Liu}}{{Liu} et~al.}{2013}]{Liu_2013b}
{Liu} G.,  {Zakamska} N.~L.,  {Greene} J.~E.,  {Nesvadba} N.~P.~H.,   {Liu} X.,
   2013, \mn@doi [\mnras] {10.1093/mnras/stt1755}, \href
  {http://adsabs.harvard.edu/abs/2013MNRAS.436.2576L} {436, 2576}

\bibitem[\protect\citeauthoryear{{Martin}}{{Martin}}{2005}]{Martin_2005}
{Martin} C.~L.,  2005, \mn@doi [\apj] {10.1086/427277}, \href
  {http://adsabs.harvard.edu/abs/2005ApJ...621..227M} {621, 227}

\bibitem[\protect\citeauthoryear{{Morton}}{{Morton}}{1991}]{Morton_1991}
{Morton} D.~C.,  1991, \mn@doi [\apjs] {10.1086/191601}, \href
  {http://adsabs.harvard.edu/abs/1991ApJS...77..119M} {77, 119}

\bibitem[\protect\citeauthoryear{{M{\"u}ller-S{\'a}nchez}, {Prieto}, {Hicks},
  {Vives-Arias}, {Davies}, {Malkan}, {Tacconi}  \&
  {Genzel}}{{M{\"u}ller-S{\'a}nchez} et~al.}{2011}]{Muller-Sanchez_2011}
{M{\"u}ller-S{\'a}nchez} F.,  {Prieto} M.~A.,  {Hicks} E.~K.~S.,  {Vives-Arias}
  H.,  {Davies} R.~I.,  {Malkan} M.,  {Tacconi} L.~J.,   {Genzel} R.,  2011,
  \mn@doi [\apj] {10.1088/0004-637X/739/2/69}, \href
  {http://adsabs.harvard.edu/abs/2011ApJ...739...69M} {739, 69}

\bibitem[\protect\citeauthoryear{{Noordermeer}, {Merrifield}  \&
  {Arag{\'o}n-Salamanca}}{{Noordermeer} et~al.}{2008}]{Noordermeer_2008}
{Noordermeer} E.,  {Merrifield} M.~R.,   {Arag{\'o}n-Salamanca} A.,  2008,
  \mn@doi [\mnras] {10.1111/j.1365-2966.2008.13487.x}, \href
  {http://adsabs.harvard.edu/abs/2008MNRAS.388.1381N} {388, 1381}

\bibitem[\protect\citeauthoryear{{Osterbrock}}{{Osterbrock}}{1989}]{Osterbrock_1989}
{Osterbrock} D.~E.,  1989, {Astrophysics of gaseous nebulae and active galactic
  nuclei}

\bibitem[\protect\citeauthoryear{{Reyes} et~al.,}{{Reyes}
  et~al.}{2008}]{Reyes_2008}
{Reyes} R.,  et~al., 2008, \mn@doi [\aj] {10.1088/0004-6256/136/6/2373}, \href
  {http://adsabs.harvard.edu/abs/2008AJ....136.2373R} {136, 2373}

\bibitem[\protect\citeauthoryear{{Riffel} \& {Storchi-Bergmann}}{{Riffel} \&
  {Storchi-Bergmann}}{2011}]{Riffel_2011}
{Riffel} R.~A.,  {Storchi-Bergmann} T.,  2011, \mn@doi [\mnras]
  {10.1111/j.1365-2966.2011.19441.x}, \href
  {http://adsabs.harvard.edu/abs/2011MNRAS.417.2752R} {417, 2752}

\bibitem[\protect\citeauthoryear{{Singh} et~al.,}{{Singh}
  et~al.}{2013}]{Singh_2013}
{Singh} R.,  et~al., 2013, \mn@doi [\aap] {10.1051/0004-6361/201322062}, \href
  {http://adsabs.harvard.edu/abs/2013A%26A...558A..43S} {558, A43}

\bibitem[\protect\citeauthoryear{{Smee} et~al.,}{{Smee}
  et~al.}{2013}]{Smee_2013}
{Smee} S.~A.,  et~al., 2013, \mn@doi [\aj] {10.1088/0004-6256/146/2/32}, \href
  {http://adsabs.harvard.edu/abs/2013AJ....146...32S} {146, 32}

\bibitem[\protect\citeauthoryear{{Somerville}, {Hopkins}, {Cox}, {Robertson}
  \& {Hernquist}}{{Somerville} et~al.}{2008}]{Somerville_2008}
{Somerville} R.~S.,  {Hopkins} P.~F.,  {Cox} T.~J.,  {Robertson} B.~E.,
  {Hernquist} L.,  2008, \mn@doi [\mnras] {10.1111/j.1365-2966.2008.13805.x},
  \href {http://adsabs.harvard.edu/abs/2008MNRAS.391..481S} {391, 481}

\bibitem[\protect\citeauthoryear{{Soto} \& {Martin}}{{Soto} \&
  {Martin}}{2012}]{Soto_2012}
{Soto} K.~T.,  {Martin} C.~L.,  2012, \mn@doi [\apjs]
  {10.1088/0067-0049/203/1/3}, \href
  {http://adsabs.harvard.edu/abs/2012ApJS..203....3S} {203, 3}

\bibitem[\protect\citeauthoryear{{Soto}, {Martin}, {Prescott}  \&
  {Armus}}{{Soto} et~al.}{2012}]{Soto_2012b}
{Soto} K.~T.,  {Martin} C.~L.,  {Prescott} M.~K.~M.,   {Armus} L.,  2012,
  \mn@doi [\apj] {10.1088/0004-637X/757/1/86}, \href
  {http://adsabs.harvard.edu/abs/2012ApJ...757...86S} {757, 86}

\bibitem[\protect\citeauthoryear{{Stern}, {Faucher-Gigu{\`e}re}, {Zakamska}  \&
  {Hennawi}}{{Stern} et~al.}{2016}]{Stern_2016}
{Stern} J.,  {Faucher-Gigu{\`e}re} C.-A.,  {Zakamska} N.~L.,   {Hennawi} J.~F.,
   2016, \mn@doi [\apj] {10.3847/0004-637X/819/2/130}, \href
  {http://adsabs.harvard.edu/abs/2016ApJ...819..130S} {819, 130}

\bibitem[\protect\citeauthoryear{{Storchi-Bergmann}, {McGregor}, {Riffel},
  {Sim{\~o}es Lopes}, {Beck}  \& {Dopita}}{{Storchi-Bergmann}
  et~al.}{2009}]{Storchi-Bergmann_2009}
{Storchi-Bergmann} T.,  {McGregor} P.~J.,  {Riffel} R.~A.,  {Sim{\~o}es Lopes}
  R.,  {Beck} T.,   {Dopita} M.,  2009, \mn@doi [\mnras]
  {10.1111/j.1365-2966.2009.14388.x}, \href
  {http://adsabs.harvard.edu/abs/2009MNRAS.394.1148S} {394, 1148}

\bibitem[\protect\citeauthoryear{{Storchi-Bergmann}, {Lopes}, {McGregor},
  {Riffel}, {Beck}  \& {Martini}}{{Storchi-Bergmann}
  et~al.}{2010}]{Storchi_2010}
{Storchi-Bergmann} T.,  {Lopes} R.~D.~S.,  {McGregor} P.~J.,  {Riffel} R.~A.,
  {Beck} T.,   {Martini} P.,  2010, \mn@doi [\mnras]
  {10.1111/j.1365-2966.2009.15962.x}, \href
  {http://adsabs.harvard.edu/abs/2010MNRAS.402..819S} {402, 819}

\bibitem[\protect\citeauthoryear{{Sun}, {Greene}, {Zakamska}  \&
  {Nesvadba}}{{Sun} et~al.}{2014}]{Sun_2014}
{Sun} A.-L.,  {Greene} J.~E.,  {Zakamska} N.~L.,   {Nesvadba} N.~P.~H.,  2014,
  \mn@doi [\apj] {10.1088/0004-637X/790/2/160}, \href
  {http://adsabs.harvard.edu/abs/2014ApJ...790..160S} {790, 160}

\bibitem[\protect\citeauthoryear{{Veilleux} \& {Osterbrock}}{{Veilleux} \&
  {Osterbrock}}{1987}]{Veilleux_1987}
{Veilleux} S.,  {Osterbrock} D.~E.,  1987, \mn@doi [\apjs] {10.1086/191166},
  \href {http://adsabs.harvard.edu/abs/1987ApJS...63..295V} {63, 295}

\bibitem[\protect\citeauthoryear{{Veilleux} et~al.,}{{Veilleux}
  et~al.}{2013}]{Veilleux_2013}
{Veilleux} S.,  et~al., 2013, \mn@doi [\apj] {10.1088/0004-637X/776/1/27},
  \href {http://adsabs.harvard.edu/abs/2013ApJ...776...27V} {776, 27}

\bibitem[\protect\citeauthoryear{{Westfall}}{{Westfall}}{2017}]{Westfall_2017}
{Westfall} K.~B.,  2017, in prep.

\bibitem[\protect\citeauthoryear{{Wild}, {Heckman}  \& {Charlot}}{{Wild}
  et~al.}{2010}]{Wild_2010}
{Wild} V.,  {Heckman} T.,   {Charlot} S.,  2010, \mn@doi [\mnras]
  {10.1111/j.1365-2966.2010.16536.x}, \href
  {http://adsabs.harvard.edu/abs/2010MNRAS.405..933W} {405, 933}

\bibitem[\protect\citeauthoryear{{Wright} et~al.,}{{Wright}
  et~al.}{2010}]{Wright_2010}
{Wright} E.~L.,  et~al., 2010, \mn@doi [\aj] {10.1088/0004-6256/140/6/1868},
  \href {http://adsabs.harvard.edu/abs/2010AJ....140.1868W} {140, 1868}

\bibitem[\protect\citeauthoryear{{Wu}, {Cao}, {Hao}, {Liu}, {Wang}, {Xia},
  {Deng}  \& {Young}}{{Wu} et~al.}{2005}]{Wu_2005}
{Wu} H.,  {Cao} C.,  {Hao} C.-N.,  {Liu} F.-S.,  {Wang} J.-L.,  {Xia} X.-Y.,
  {Deng} Z.-G.,   {Young} C.~K.-S.,  2005, \mn@doi [\apjl] {10.1086/497961},
  \href {http://adsabs.harvard.edu/abs/2005ApJ...632L..79W} {632, L79}

\bibitem[\protect\citeauthoryear{{Wylezalek}}{{Wylezalek}}{2017}]{Wylezalek_2017}
{Wylezalek} D.,  2017, in prep.

\bibitem[\protect\citeauthoryear{{Wylezalek} \& {Zakamska}}{{Wylezalek} \&
  {Zakamska}}{2016}]{Wylezalek_2016b}
{Wylezalek} D.,  {Zakamska} N.~L.,  2016, \mn@doi [\mnras]
  {10.1093/mnras/stw1557}, \href
  {http://adsabs.harvard.edu/abs/2016MNRAS.461.3724W} {461, 3724}

\bibitem[\protect\citeauthoryear{{Yan} et~al.,}{{Yan} et~al.}{2016}]{Yan_2016}
{Yan} R.,  et~al., 2016, preprint, \href
  {http://adsabs.harvard.edu/abs/2016arXiv160708613Y} {} (\mn@eprint {arXiv}
  {1607.08613})

\bibitem[\protect\citeauthoryear{{Zakamska} \& {Greene}}{{Zakamska} \&
  {Greene}}{2014}]{Zakamska_2014}
{Zakamska} N.~L.,  {Greene} J.~E.,  2014, \mn@doi [\mnras]
  {10.1093/mnras/stu842}, \href
  {http://adsabs.harvard.edu/abs/2014MNRAS.442..784Z} {442, 784}

\bibitem[\protect\citeauthoryear{{Zakamska} et~al.,}{{Zakamska}
  et~al.}{2016}]{Zakamska_2016}
{Zakamska} N.~L.,  et~al., 2016, \mn@doi [\mnras] {10.1093/mnras/stv2571},
  \href {http://adsabs.harvard.edu/abs/2016MNRAS.455.4191Z} {455, 4191}

\bibitem[\protect\citeauthoryear{{Zubovas} \& {King}}{{Zubovas} \&
  {King}}{2012}]{Zubovas_2012}
{Zubovas} K.,  {King} A.~R.,  2012, \mn@doi [\mnras]
  {10.1111/j.1365-2966.2012.21845.x}, \href
  {http://adsabs.harvard.edu/abs/2012MNRAS.426.2751Z} {426, 2751}

\bibitem[\protect\citeauthoryear{{van der Kruit} \& {Freeman}}{{van der Kruit}
  \& {Freeman}}{1986}]{van_der_Kruit_1986}
{van der Kruit} P.~C.,  {Freeman} K.~C.,  1986, \mn@doi [\apj]
  {10.1086/164102}, \href {http://adsabs.harvard.edu/abs/1986ApJ...303..556V}
  {303, 556}

\makeatother
\end{thebibliography}





\bsp	
\label{lastpage}
\end{document}